\pdfoutput=1
\documentclass[12pt]{article}

\usepackage[right=1.25in,left=1.25in,top=1.1in,bottom=1.1in]{geometry}

\usepackage[T1]{fontenc}
\usepackage[utf8]{inputenc}
\usepackage[english]{babel}

\usepackage{amsmath,amsthm,amsfonts,amssymb,bm}

\usepackage{graphicx}
\usepackage[section]{placeins}
\usepackage{caption}
\usepackage{subcaption}   %
\usepackage{threeparttable}
\usepackage{multirow}
\usepackage{tabularx,ragged2e,booktabs}
\usepackage{array}

\usepackage[natbibapa]{apacite}
\setcitestyle{aysep={}}

\usepackage{color}
\usepackage{setspace}
\usepackage[export]{adjustbox}
\usepackage{titlesec}
\usepackage[title]{appendix}
\usepackage{ulem}      %
\usepackage{sectsty}
\usepackage{comment}
\usepackage{footmisc}
\usepackage{pdflscape}

\usepackage{hyperref}
\hypersetup{
  colorlinks,
  citecolor=blue,
  filecolor=blue,
  linkcolor=blue,
  urlcolor=blue
}

\definecolor{darkgreen}{rgb}{0.0,0.7,0.0}

\newtheorem{proposition}{Proposition} 
 
\newtheorem{theorem}{Theorem}

\newenvironment{keywords}{\par\noindent\textbf{Keywords:}~}{\par}

\newcolumntype{L}[1]{>{\raggedright\let\newline\\arraybackslash\hspace{0pt}}m{#1}}
\newcolumntype{C}[1]{>{\centering\let\newline\\arraybackslash\hspace{0pt}}m{#1}}
\newcolumntype{R}[1]{>{\raggedleft\let\newline\\arraybackslash\hspace{0pt}}m{#1}}

\begin{document}

\title{Joint Calibration of the Volatility Surface and Variance Term Structure\thanks{Code for this article is available at \url{https://github.com/jiwookwhy/Joint_calibration}.}}

\author{Jiwook Yoo\thanks{I thank David Hsieh, Douglas Breeden, Paymon Khorrami and Ravi Bansal for their feedback on many drafts of this article, and the finance faculty at the Fuqua School of Business for their helpful comments. The views expressed and any errors are my own. Email: \texttt{jiwook.yoo@duke.edu}.}\\
\small Fuqua School of Business (Duke University), Durham, NC, USA}

\maketitle

\begin{abstract}
\singlespacing
\noindent I study the calibration of option pricing models with an objective that combines errors on the implied volatility surface and on the variance term structure (VTS), and how the weight on the two trades one fit against the other. I then ask how the calibration affects hedges built from the fitted model. The setting is the stochastic volatility jump-diffusion model of Bates (1996), calibrated to S\&P~500 index options on each trading day from 1996 to 2025. Calibrated to option prices alone, the model misses the VTS; however, a small weight on the VTS sharply reduces the error at the cost of a small loss in accuracy on the volatility surface and little change in hedging performance. I fit the model in two ways: unrestricted, with every parameter free, and restricted, with the spot variance pinned to an estimate from high-frequency returns and one jump-size law for all dates, estimated from option prices pooled across dates. This choice matters more for hedging than the weight does, and the closer fit to option prices does not necessarily have the lower hedging errors. The unrestricted fits match the surface more closely, but with rare and implausibly large jumps. Once options enter the quadratic hedge, whose positions depend on the whole fitted jump law, the restricted fits' hedging errors are as low or lower. The penalty on the VTS reduces the extreme jump sizes implied by the fitted parameters, but it does not explicitly exclude such jump laws.
\end{abstract}

\begin{singlespace}\raggedright
\end{singlespace}
\clearpage

\section{Introduction}
Parametric option pricing models are workhorses of contemporary academic research and industry practice. Broadly speaking, selecting a model amounts to specifying the stochastic process for the underlying asset's price, and if applicable, other relevant state variables, with the parameters of these processes often calibrated to closely match observed market prices. The distribution implied by these processes in turn informs a risk-neutral measure from which derivative prices and hedge ratios are derived. This workflow has its roots in the classic works of \cite{black1973pricing} and \cite{black1976pricing}, who model the underlying asset as a geometric Brownian motion (GBM). However, a large body of evidence suggests that asset prices exhibit time series properties that differ from a simple GBM. Asset prices can jump \citep{carr2002fine, bollerslev2011tails}, possess time-varying volatility \citep{shephard2005stochastic}, exhibit a leverage effect \citep{carr2017leverage}, and deviate in a myriad of other ways not captured by a simple drift-diffusion process. 

Researchers and industry practitioners alike have considered more complex models which aim to shore up deficiencies by more faithfully modeling these underlying processes. However, these more sophisticated models are more difficult to calibrate, and it is well known that calibrating them, by minimizing the distance between model and market prices, can be an ill-posed problem \citep{cont2004nonparametric}. For instance, the objective function may not be globally convex or may possess very flat regions; in such cases, the numerical optimization procedure used to minimize the objective function may fail to converge to a minimum. A large class of approaches therefore supplements the standard objective, a sum of squared errors over the volatility surface, with additional terms as a means of regularization \citep{cont2003financial}.

This article examines model calibration and its consequences for hedging performance when the objective function includes a penalty for deviations of model-implied variances from those observed in the market. This is not the first paper to consider a variance-based penalty in model calibration, nor the first to ask whether the calibration method matters for hedging \citep{alexander2012does}. To my knowledge, however, it is the first to vary the weight on a penalty for the variance term structure (VTS) and to trace how that weight trades off the fit to option prices against the fit to the VTS. I also compare how calibration choices regarding both the penalty weight and restrictions that discipline the variance and jump process affect hedging performance. To fix ideas, consider out-of-the-money option prices on a common underlying with maturities $\tau_1, \ldots, \tau_M$ and $N(\tau_j)$ options at maturity $\tau_j$.\footnote{I treat an option with strike $K$ as out-of-the-money if $K \geq F$ for a call and $K < F$ for a put, where $F$ is the forward price of the underlying for the option's maturity.} I consider objective functions of the form:
\begin{equation}\label{eq:obj_function_intro}
\begin{aligned}
&\alpha   \sum\limits_{j=1}^{M}\sum\limits_{i=1}^{N(\tau_j)} \left[\sigma^{mkt}(K_{i,j}, \tau_j) - \sigma^{mod}(K_{i,j}, \tau_j\vert \Theta) \right]^2\\
&\qquad + (1-\alpha) \sum\limits_{j=1}^{M} N(\tau_j) \left[\sqrt{VTS^{mkt}(\tau_j)} - \sqrt{VTS^{mod}(\tau_j\vert \Theta)} \right]^2
\end{aligned}
\end{equation}
where $\sigma^{mkt}(K_{i,j}, \tau_j)$ denotes the Black--Scholes implied volatility (BSIV) of an OTM European-style option with strike price $K_{i,j}$ and maturity $\tau_j$, which I obtain by inverting the option's observed price with the \cite{black1976pricing} formula on the forward price.\footnote{With the dividend yield absorbed into the forward price, this is the \cite{black1973pricing} implied volatility.} Likewise, $\sigma^{mod}(K_{i,j}, \tau_j\vert \Theta)$ is the BSIV obtained from inverting the model price of an OTM option given the vector of the model parameters $\Theta$. Building off this notation, I use $VTS^{mkt}(\tau)$ and $VTS^{mod}(\tau\vert\Theta)$ to denote the variance term structure of the underlying observed in the market and implied by the model parameterized by $\Theta$ respectively.\footnote{Following the conventions for quoting variance derivatives, the VTS consists of annualized variances, and \eqref{eq:obj_function_intro} compares their square roots, in volatility units.} Finally, $\alpha \in [0,1]$ is a hyperparameter that governs the strength of the penalty term, in this case, model-implied deviations from the observed VTS. As $\alpha$ lies in the unit interval, \eqref{eq:obj_function_intro} can also be interpreted as the weighted average between volatility surface and VTS errors. 

The objective function above nests two special cases at the boundary. When $\alpha$ equals $1$, it reduces to the standard sum of squared errors in implied volatility (IV) space, one of the most common objective functions used in the literature to calibrate option pricing models. When $\alpha$ is zero, it consists purely of errors along the VTS. I apply it to the stochastic volatility jump-diffusion model of \cite{bates1996jumps}, fitted to S\&P~500 index options on each trading day from 1996 to 2025, in two forms. The \emph{unrestricted} fits estimate all eight parameters every day. The \emph{restricted} fits take the parameters that option prices identify least well from information outside the day's option cross-section: the spot diffusive variance is pinned to a high-frequency estimate, which imposes as a constraint the restriction that \cite{andersen2015parametric} impose through a penalty, and a single jump-size law is shared across all dates, in the spirit of \cite{bates2000post}.\footnote{As \cite{broadie2007model} note, option prices identify the product of the jump intensity and the mean jump size, but not its two components separately; \cite{pan2002jump} raises the same concern about separating the risk premia for jump timing and jump size. Often, some jump parameters are therefore fixed using physical-measure information. \cite{pan2002jump}, \cite{eraker2004stock} and \cite{broadie2007model} set the risk-neutral jump intensity equal to the physical one, which \cite{broadie2007model} take from a returns-based estimate, and \cite{bardgett2019inferring} do the same and also fix the physical jump-size moments at estimates from high-frequency returns. I instead share one jump-size law across all dates, estimated from option prices pooled across dates, and allow the intensity to vary from day to day.} The restricted fits are the calibration this article proposes, because they restore a plausible jump law. The unrestricted fits serve as the comparison: at $\alpha = 1$, they are the straightforward and conventional practice of minimizing squared implied volatility errors with every parameter free.

The fits deliver three findings. First, option prices by themselves do not pin down the VTS, even though the VTS is computed from those same prices, and the error can be corrected at the cost of a small loss in accuracy on the volatility surface. Calibrated the conventional way, at $\alpha = 1$ with every parameter free, the model matches the surface to within 0.6 volatility points on average but misses the VTS by 1.3 points. The errors are largest at maturities of at most nine days, in crises and in recent years, when short-dated options are more prolifically listed. A small weight on the VTS sharply reduces the error: the VTS error is 0.8 points at $\alpha = 0.9$ and 0.5 points at $\alpha = 1/2$, while the surface error rises only from 0.6 to 0.7 points. Second, the daily fit does not pin down the jump law either. Left unrestricted, the model prices options with rare and very large jumps: at the median, 0.06 jumps a year with a mean of $-16\%$ and a volatility of 26\%, and on three days in ten a mean jump larger than 25\% in absolute value. Restricting the fit gives about one jump a year, with a mean of $-4\%$ and a volatility of 8\%, and increases the surface fit errors by about half a volatility point. Third, the two forms of discipline overlap but are not substitutes. The restrictions lower the VTS errors without any penalty, and the penalty reduces the extreme jump sizes implied by the fitted parameters, but it does not explicitly exclude such jump laws.

These differences matter for hedging, because a hedge built from a fitted model depends on how its jump law spreads jump risk across strikes. I hedge options at five points of the smile in three expiries with seven hedges, which I rebalance daily and weekly. The hedges range from the practitioner's Black--Scholes and model deltas and delta--vega hedges to the quadratic hedges of \cite{cont2007hedging}, which minimize the risk-neutral variance of the hedging error with the index and up to two options. The main finding is that the closer fit to option prices does not necessarily have the lower hedging errors. Which fit has the lower hedging errors depends on which instruments are available for inclusion in the hedge. With the index alone, and with the model delta--vega hedge, the unrestricted fits have lower hedging errors. Once options enter the quadratic hedge, the restricted fits' hedging errors are as low or lower, at every value of $\alpha$ and at both horizons. With one option, their advantage is significant under the conventional calibration, at $\alpha = 1$, and it narrows as the weight on the VTS rises, which is consistent with the penalty reducing the extreme jump sizes implied by the unrestricted fits. With two options, it is significant at every positive value of $\alpha$. The second option lowers the effectiveness of both fits' hedges, and of the unrestricted fits' by far more. A calibration can therefore be chosen for a plausible jump law, at some cost in fit, and still hedge well with options. For the weight itself, values of $\alpha$ between $1/2$ and 0.9 reduce the VTS error at little cost on the surface, sharply so for the unrestricted fits, and hedging performance changes little within that range.

Finally, none of the model hedges improves on the practitioner's Black--Scholes delta--vega hedge on average, as \cite{alexander2012does} also find for the \cite{heston1993closed} model; the quadratic hedges with one option match it only in periods with jumps. That comparison is not like for like. The Black--Scholes model has a single volatility parameter for the diffusion of the index, so as a pricing model it values every option with the same volatility. The practitioner's hedge instead takes each option's Greeks at that option's own implied volatility, as if the index followed a different diffusion for each option, so its hedge ratios rest on inconsistent assumptions about the price process.

This article makes three contributions to the literature reviewed in Section~\ref{subsec: related_lit}. First, I show that, in a model with jumps, the calibration that fits option prices most closely does not necessarily have the lowest hedging errors once options are part of the hedging portfolio. I compare unrestricted daily fits with fits whose diffusive variance and jump sizes are disciplined by outside information. That option prices identify jump laws poorly is well known \citep{pan2002jump, broadie2007model, he2006calibration}, and restrictions of this kind have become increasingly popular. The contribution comes from examining their consequences for hedging, using quadratic hedges that account for jump risk and use options as hedges, at two horizons and across different relative weightings of the volatility surface and VTS. This qualifies a finding of \cite{alexander2012does} that, for the \cite{heston1993closed} model, a better fit to option prices delivers lower hedging errors. Their model has no jumps, their calibrations place no penalty on the VTS, and the hedges for which they find this are minimum-variance and delta--vega hedges, none of which accounts for jump risk. In a model with jumps, the finding holds for hedges with the underlying alone and for delta--vega hedges, but not once options enter the quadratic hedge. Second, I treat the weight on the VTS as a hyperparameter and vary it. \cite{nagashima2014asymptotic} fix the balance between option prices and variance swap rates, and \cite{andersen2017short} derive the weight on their penalty, which is on the spot variance rather than the VTS, from estimates of measurement error. I show that the VTS errors of the conventional calibration can be corrected at the cost of a small loss in accuracy on the volatility surface. Third, I show how the two forms of discipline interact across values of $\alpha$: the restrictions lower the VTS errors much as the penalty does, and the penalty narrows the gap between the two fits' one-option quadratic hedges, but it does not explicitly exclude the extreme jump laws of the unrestricted fits. 

The remainder of this article is organized as follows. Section~\ref{subsec: related_lit} reviews related approaches from the literature. Section~\ref{sec: theory} sets out the theory, the calibration framework and the hedges I evaluate, including the quadratic hedge of \cite{cont2007hedging} in the Bates model. Section~\ref{sec: data} describes the panel of S\&P~500 options and variance term structures. Section~\ref{subsec: model_calibration} compares the unrestricted and restricted fits of the Bates model and how $\alpha$ trades the fit to the surface against the fit to the VTS. Section~\ref{subsec: hedging} compares their hedging performance at daily and weekly horizons and across values of $\alpha$. Section~\ref{sec: conclusion} concludes. Appendix~\ref{app: proofs} proves Propositions~\ref{prop: bates_vts} and~\ref{prop: bates_hedge}, and Appendix~\ref{app: pooled} describes the estimation of the pooled jump-size law.

\subsection{Related literature: Calibration approaches}\label{subsec: related_lit}
In both financial practice and the academic literature, the calibration of parametric option pricing models is routine. Systematic approaches to model calibration involve searching for parameters that best fit observed market prices or implied volatilities by some criterion. Early work, such as the study of \cite{amin1994implied} on Eurodollar options and futures, minimized the sum of squared errors in option prices to calibrate the interest rate model of \cite{heath1992bond}. \cite{bakshi1997empirical} likewise minimize the sum of squared errors in option prices. Another popular alternative involves approximating the difference in implied volatilities using option prices and \cite{black1973pricing} vegas \citep[e.g.,][]{christoffersen2009shape}.\footnote{This is the well-known ``vega-weighting'' approximation to the sum of squared IV errors in \eqref{eq:obj_function_intro}. It substitutes a first-order Taylor polynomial in the BSIV to approximate $\sigma^{mod}(K_{i,j}, \tau_j\vert \Theta) - \sigma^{mkt}(K_{i,j}, \tau_j)$ for each contract.}

Related to the standard approach of minimizing squared errors is the more general approach of using moment conditions under the generalized method of moments framework introduced by \cite{hansen1982large}. \cite{chernov2000study} estimate the \cite{heston1993closed} model by the efficient method of moments from index returns, from at-the-money implied volatilities, and from both jointly; out of sample, the estimates from options alone price best, and hedging performance is insensitive to which data are used. \cite{pan2002jump} expands on this approach, estimating a continuous-time jump-diffusion model using moments from the physical measure and option-implied moments to recover risk premia associated with jumps. \cite{broadie2007model} instead fix the parameters that must agree under the two measures at estimates from earlier returns-based studies, estimate the risk premia from option prices, and investigate the effects of model misspecification on inferred volatility and jump risk premia.

Another class of approaches, which includes my own proposed methodology, adds a regularization term to the usual sum of squared errors in BSIVs. The general form of such a regularized objective function is given in \eqref{eq: obj_reg_ivs}:
\begin{equation}\label{eq: obj_reg_ivs}
       \left( \sum\limits_{j=1}^M   \sum\limits_{i=1}^{N(\tau_j)} w_{i,j}\left[\sigma^{mod}(K_{i,j}, \tau_j\vert \Theta) - \sigma^{mkt}(K_{i,j}, \tau_j)\right]^2 \right) + h(\Lambda;\Theta)
\end{equation}
where $w_{i,j}$ are non-negative weights and $h(\Lambda;\Theta)$ the regularization term given regularization hyperparameters $\Lambda$ and model parameters $\Theta$. As discussed in detail by \cite{cont2003financial}, the inverse problem of calibration can be ill-posed or susceptible to issues related to numerical optimization. Intuitively, regularization can modify the objective function to endow it with desirable mathematical properties (e.g., convexity, additional curvature) which are more amenable to numerical optimization. \cite{cont2004nonparametric} propose a regularization term $h$ based on relative entropy. Specifically, the authors suggest setting the penalization term $h(c; \Theta)$ in \eqref{eq: obj_reg_ivs} to $cD(\mathbb{Q}_\Theta, \mathbb{P}_0)$ where $D$ is the Kullback-Leibler (KL) divergence of the model-implied pricing measure $\mathbb{Q}_\Theta$ with respect to some prior measure $\mathbb{P}_0$.\footnote{In this context, $c$ is a scalar hyperparameter that regulates the relative importance of the KL penalty and the sum of squared errors.} \cite{lagnado1997technique} calibrate a local volatility function $\sigma$ by adding to the sum of squared pricing errors the penalty $\beta\Vert\nabla\sigma\Vert_2^2$, the squared $L^2$ norm of the function's gradient weighted by $\beta > 0$. This Tikhonov-type penalty favors smooth volatility surfaces and stabilizes an otherwise ill-posed inverse problem. For the same problem, \cite{egger2005tikhonov} demonstrate stable convergence of Tikhonov regularization and provide recommendations for the choice of norm motivated by theoretical results. Another common Tikhonov regularization term is a penalty based on the Euclidean distance $\Vert\Theta - \Theta_0\Vert^2$ from a pre-selected parameter vector $\Theta_0$ as in \cite{nagashima2014asymptotic}.

Approaches most similar to the one I examine here are variations of \eqref{eq: obj_reg_ivs} in which the regularization term $h$ penalizes deviations of a model-implied measure of variance from an observed one. \cite{andersen2015parametric} augment the sum of squared errors in BSIVs with a penalty on the distance between the model's spot diffusive variance and a nonparametric estimate of diffusive variance from high-frequency returns. The penalty rests on a no-arbitrage restriction, namely that diffusive variance is the same under the physical and risk-neutral measures; its weight is small and fixed, and serves to regularize the optimization objective and keep spot volatility in a sensible range. The same criterion underlies \cite{andersen2015risk}, \cite{andersen2017short} and \cite{andersen2020pricing}, with \cite{andersen2017short} deriving the weight on the penalty term from estimates of the measurement error in the option prices and in the high-frequency estimate, rather than fixing it. \cite{bardgett2019inferring} maximize a joint likelihood of index returns, S\&P~500 and VIX options, and the VIX index, a 30-day risk-neutral measure of variance. The closest precedent for my objective is \cite{nagashima2014asymptotic}, who calibrate a two-factor Heston model in two steps. The first step fits the term structure of variance swap rates, with maturities from one to 24 months, and the second fixes the initial variances at their first-step values and fits the remaining parameters to vega-weighted option prices and the variance swap rates together. Both steps add a Tikhonov penalty on the model parameters, and the variance swap rates enter in volatility units. Up to that penalty, their second step is the case $\alpha = 1/2$ of the objective I study, with the variance swap rates weighted equally rather than by the number of options at each maturity, but the balance between the option and variance terms is fixed by construction and never varied. 

None of these papers studies the consequences of regularization on hedging. Two papers do hedge after a calibration that utilizes variance information. \cite{pacati2018smiling} calibrate to SPX options, the VIX futures curve and VIX options with equal weights on the model errors in each and consider option hedges using the index and two VIX futures. \cite{alexander2012does} ask directly whether the calibration method matters for hedging. For the \cite{heston1993closed} model on FTSE~100 options, they compare daily calibration to option prices, a pre-sample calibration in which a 30-day implied volatility index pins down the spot variance, and parameters estimated from index returns. With the minimum-variance and delta--vega hedges they focus on, daily recalibration gives the lowest hedging errors and the returns-based parameters the highest. They also note that the empirical hedging literature finds little evidence that including jumps improves hedging performance. Their setting differs from this article's in three ways. First, their model has no jumps. Their minimum-variance hedges therefore minimize the variance of the hedging error against diffusive moves only, and none of their hedges, including those with an option, accounts for jump risk. The quadratic hedges here minimize the variance of the hedging error including jumps, under each fit's jump law. Second, the volatility index enters as a constraint on the spot variance rather than as a penalty term in the objective. Third, each of their hedges uses at most one option, whereas the quadratic hedges here use up to two.

\cite{he2006calibration} study calibration and hedging together for a jump-diffusion model with a local volatility function, on option prices simulated from known parameters. They find that the parameters of the jump-size distribution are difficult to estimate from vanilla option prices, that the calibration is well posed once that distribution is given, and that hedges with the underlying and several options are not much affected by the poor estimates. The restricted fits here follow their second observation, fixing the jump-size law and leaving the intensity free. My setting differs from theirs in two ways: I use market rather than simulated prices, and my hedges use at most two options. In this setting, two jump laws that fit option prices about equally well give different hedging performance.

\section{Theory and calibration framework}\label{sec: theory}
\subsection{Identification of the variance term structure}\label{subsec: vs}
This subsection defines the variance term structure and explains why the squared VIX at each maturity, which option prices determine, can stand in for the variance swap rate. I work in continuous time, on a filtered probability space with the usual conditions \citep{protter2004stochastic}, and take all prices and expectations at time 0 unless noted otherwise. The underlying $S_t$ has risk-neutral dynamics with a diffusive term and a jump component that arrives according to a Poisson process: 
\begin{equation}\label{eq: general_process}
d S_t=\left(r-q-\nu_J\right) S_{t-} d t+\sqrt{V_t} S_{t-} d W_t+\left[\exp(J_t)-1\right] S_{t-} d N_t, \qquad  \mathbb{P}^{\mathbb{Q}}(dN_t = 1) = \lambda\, dt
\end{equation}
where $\mathbb{P}^{\mathbb{Q}}$ denotes probability under the risk-neutral measure $\mathbb{Q}$, $r$ is the risk-free rate, $q$ the dividend yield, and $\nu_J = \lambda\mathbb{E}^{\mathbb{Q}}\left[e^{J_t}-1\right]$ compensates the jump component, so that the price drifts at $r-q$ under $\mathbb{Q}$. $S_{t-}$ is the price at time $t$ prior to the arrival of a jump. Whenever the meaning of an expression is unambiguous, I omit time subscripts. I impose the following conditions on the process: (1) the variance process $V_t$ has finite long-run mean, and (2) the log jump size $J_t$ has a finite second moment and $\mathbb{E}^{\mathbb{Q}}\left[e^{J_t}\right] < \infty$. The stochastic process in equation \eqref{eq: general_process} nests many familiar price processes featured in parametric option pricing models. Examples include the \cite{black1973pricing} model, the \cite{heston1993closed} stochastic volatility model, and the \cite{bates1996jumps} model, which pairs stochastic volatility with the jump process from the \cite{merton1976option} model.

In continuous time, the realized variance of a price process is measured by the quadratic variation of its log price. For the price process in \eqref{eq: general_process}, the quadratic variation from time 0 to $\tau$, $QV(\tau)$, is given by:
\begin{equation}\label{eq: QV}
 QV(\tau) = \int_0^{\tau} \left(d\log S_t\right)^2 = \int_0^{\tau}  V_t\, d t + \sum_{k=1}^{N_\tau} J_k^2,
\end{equation}
where $N_\tau$, the Poisson process of \eqref{eq: general_process} at time $\tau$, counts the jumps in $[0,\tau]$ and $J_k$ is the $k$-th log jump size. Derivatives on quadratic variation, most notably variance swaps and variance futures, pay a function of annualized quadratic variation.\footnote{In practice, the payoff of these derivatives discretizes \eqref{eq: QV}, approximating the quadratic variation by the sum of squared log returns. See \cite{bondarenko2014variance} for a detailed overview and analysis of this discretized payoff. } Absent arbitrage, the fair strike of a variance swap (VS) equals the expected quadratic variation under a pricing measure $\mathbb{Q}$.\footnote{Variance swaps allow market participants to hedge variance directly or to take positions on the market’s pricing of expected variance. The swap payoff is positive when realized variance over $[0,\tau]$ turns out to be unexpectedly higher than the variance anticipated under the pricing measure. Many modern asset pricing theories posit a negative variance risk premium in equity markets. Empirical studies lend strong support to this view, with studies finding negative average returns on traded and synthetic variance swaps \citep{carr2009variance, konstantinidi2016does}.} Market convention annualizes this expected variance. As a function of maturity $\tau$, the resulting variance swap rates $VS(\tau)$ form the VTS:
\begin{equation}\label{eq: imp_var}
 VS(\tau) = \dfrac{1}{\tau}  \mathbb{E}^{\mathbb{Q}}\left[QV(\tau) \right] = \dfrac{1}{\tau}  \mathbb{E}^{\mathbb{Q}}\left[\int_0^{\tau} V_t\,  d t\right] + \lambda\,\mathbb{E}^{\mathbb{Q}}\left[J^2\right]
\end{equation}

Throughout the 1990s, developments by \cite{dupire1993model}, \cite{neuberger1994log}, \cite{carr1998towards}, and \cite{demeterfi1999} provided the theoretical basis for the valuation of contracts that pay the expected variance as in \eqref{eq: imp_var}. The core idea is that a static portfolio of options and a dynamic position in the underlying replicate the variance payoff.\footnote{See \cite{bossu2005just} for a detailed reference on the dynamic replication strategy and its implementation. } As shown by \cite{britten2000option}, if the price process is continuous, then this construction is model-free. In \eqref{eq: general_process}, the price is continuous when the jump intensity $\lambda$ is zero. In that case, by a result of \cite{neuberger1994log}:
\begin{equation}\label{eq: var_diffusion}
VS(\tau) = \mathbb{E}^\mathbb{Q}\left[\dfrac{1}{\tau}\int_0^\tau V_t\, d t\right] = \dfrac{2}{\tau}\,\mathbb{E}^\mathbb{Q}\left[\int_0^\tau \dfrac{dS_t}{S_t} - \int_0^\tau d \log S_t\right] = \dfrac{2}{\tau}\,\mathbb{E}^\mathbb{Q}\left[-\log\left( \dfrac{S_\tau}{F_\tau}\right)\right]
\end{equation}
where the last equality uses $\mathbb{E}^\mathbb{Q}\left[\int_0^\tau dS_t/S_t\right] = (r-q)\tau$ and $F_\tau = S_0e^{(r-q)\tau}$, the $\tau$-forward price of the underlying at time 0. By \eqref{eq: var_diffusion}, the VS rate is set by the price of a claim on the log return, which depends only on the underlying's value at maturity, $S_{\tau}$, and not on its path. This log contract is a European-style claim, so the payoff-spanning results of \cite{breeden1978prices} and \cite{carr1998towards} price it with options.\footnote{\label{fn: spanning}For a function $g: (0,\infty)\to \mathbb{R}$ with continuous second derivative, the time 0 price of a European-style contingent claim paying $g(S_T)$ at time $T$ is equal to the price of a portfolio of bonds and options given by $e^{-rT}g(F) + \int_0^F g''(K)P_0(K) \, dK +  \int_F^\infty g''(K)C_0(K) \, dK $, where $F$ denotes the forward price for maturity $T$. The log contract is priced by setting $g(K) = -\log(K)$.}  With $C_0(K, \tau)$ and $P_0(K, \tau)$ the time 0 prices of European-style calls and puts with strike $K$ and maturity $\tau$, the replication gives the last equality in \eqref{eq: VS_cont}:
\begin{equation}\label{eq: VS_cont}
 VS(\tau) = \dfrac{2}{\tau}\, \mathbb{E}^\mathbb{Q}\left[- \log\dfrac{S_\tau}{F_\tau}\right] = \dfrac{2e^{r\tau}}{\tau}\left[ \int_0^{F_\tau} \dfrac{P_0(K,\tau)}{K^2} \, dK + \int_{F_\tau}^\infty \dfrac{C_0(K,\tau)}{K^2}  \, dK \right]
\end{equation}
Readers familiar with volatility indices will recognize \eqref{eq: VS_cont} as the formula underlying the squared VIX.

\cite{carr2012variance} call the constant 2 in front of the log contract in \eqref{eq: VS_cont} the ``multiplier''. It equals 2 when the price process is purely diffusive. When the underlying can jump, the multiplier can differ from 2, by an amount whose size and sign depend on the higher-order moments of the jump distribution, so the first equality in \eqref{eq: VS_cont} is then an approximation. 

When jumps are present, \cite{jiang2005model} show that the integral in \eqref{eq: VS_cont} remains close to the expected return variance. For the process in \eqref{eq: general_process}, the VS rate exceeds the integral by $2\lambda\mathbb{E}^{\mathbb{Q}}\left[1 + J + J^2/2 - e^{J}\right]$ (equation \eqref{eq: VIX_gap} below), whose leading term for small jumps is $-(\lambda/3)\,\mathbb{E}^{\mathbb{Q}}\left[J^3\right]$: the gap is third order in the jump size, and positive when jumps are negatively skewed.\footnote{\cite{jiang2005model} measure return variance with simple returns, $\int_0^\tau (dS_t/S_t)^2$. In that convention, the leading term of the gap is $2\lambda\mathbb{E}^{\mathbb{Q}}\left[J^3/3\right]$, with the opposite sign.} \cite{ait2020term} compare S\&P~500 variance swap quotes with the squared VIX computed from options at the 2-, 3- and 6-month maturities and find that the differences are small. I therefore use the squared VIX computed from options of maturity $\tau$, $VIX^2(\tau)$, as my VTS:
\begin{equation}\label{eq: VTS}
   VTS(\tau) := VIX^2(\tau) = \frac{2e^{r\tau}}{\tau}\left[\int_0^{F_\tau} \frac{P_0(K,\tau)}{K^2} \, dK + \int_{F_\tau}^\infty \frac{C_0(K,\tau)}{K^2} \, dK \right],
\end{equation}
where $r$ is the risk-free rate, so that $e^{r\tau}$ converts today's option prices $P_0$ and $C_0$ into forward values. 

For the remainder of this article, I adopt the convention that $VTS^{mod}(\tau\vert \Theta)$ is the quantity in \eqref{eq: VTS} computed using the put and call prices with maturity $\tau$ implied by an option pricing model with parameters $\Theta$. Likewise, I use $VTS^{mkt}(\tau)$ to denote the $VIX^2(\tau)$ implied by option prices observed in the market. Computing $VTS^{mod}(\tau\vert \Theta)$ from a model and its parameters is straightforward. Computing $VTS^{mkt}(\tau)$ from market prices is harder, as only a discrete set of strikes is observed, rather than a continuum on $(0,\infty)$. I follow the literature and use the method of \cite{jiang2005model}. For each trading date and listed maturity $\tau$, I start from the BSIVs of the out-of-the-money options: puts with strikes below the forward $F_\tau$ and calls with strikes at or above it. I interpolate these implied volatilities across strikes with a natural cubic spline in log-moneyness $\log(K/F_\tau)$, and hold them flat at their end values below the lowest and above the highest observed strike. I then convert the resulting smile back into put and call prices with the \cite{black1976pricing} formula and evaluate the two integrals in \eqref{eq: VTS} numerically.\footnote{I use Simpson's rule in log-strike with 2,001 nodes on each side of the forward. The grid extends 12 standard deviations, measured at the end-point volatilities, beyond the lowest and highest quoted strikes.} The interpolation reduces the discretization error of a finite strike grid, and the extrapolation fills in the tails beyond the listed strikes so as to reduce truncation error. For the model, I evaluate $VTS^{mod}(\tau\vert\Theta)$ in closed form, with the expression for the Bates model derived in Section~\ref{subsec: bates}. 

\subsection{Stochastic volatility with jumps: the \cite{bates1996jumps} model}\label{subsec: bates}
For the fits and hedges of Section~\ref{sec: emp_results}, I use the stochastic volatility jump-diffusion model of \cite{bates1996jumps}, a special case of \eqref{eq: general_process}. It combines the stochastic volatility process of \cite{heston1993closed} with the jump process of \cite{merton1976option}, and so nests two workhorse option pricing models. This subsection presents the model and derives its VTS in closed form. 

As in \eqref{eq: general_process}, $S_t$ is the underlying's price and $S_{t-}$ its level just before a jump at time $t$. The model adds a latent variance process, $V_t$, and gives the two the following risk-neutral dynamics:
\begin{equation}
\begin{aligned}
& d S_{t}=\left(r-q-\lambda \mu_J\right) S_{t-} d t+\sqrt{V_t} S_{t-} d W_t+(e^{J_t}-1) S_{t-} d N_t \\
& d V_t=\kappa\left(\theta-V_t\right) d t+\sigma_v \sqrt{V_t} d W_t^v \\
& \mathbb{E}\left[d W_t  d W_t^v\right]= \rho d t \\
& \mathbb{P}^{\mathbb{Q}}\left(d N_t=1\right)=\lambda d t\\
& J_t \sim \mathcal{N}\left(\log \left(1+\mu_J\right)-\frac{\sigma_J^2}{2}, \sigma_J^2 \right)
\end{aligned}
\end{equation}
For the Bates model, the parameter vector is $\Theta = [V_0, \kappa, \theta, \sigma_v, \rho, \lambda, \mu_J, \sigma_J]$. $V_t$ is a square-root diffusion whose instantaneous variance, $\sigma_v^2 V_t$, is proportional to its level; it mean-reverts to the long-run variance $\theta$ at speed $\kappa$. $\rho$ is the instantaneous correlation between the two Brownian motions. The log jump size $J_t$ is drawn from a Gaussian distribution with mean $\log(1+\mu_J) - \sigma^2_J/2$ and variance $\sigma_J^2$. Jumps arrive at intensity $\lambda$, and a jump changes the price instantly by $S_{t-}[\exp(J_t) - 1]$. For the remainder of this article, I use the notation $\mathbb{E}^\mathbb{Q}(\cdot \vert \Theta)$ and $VTS^{mod}(\tau\vert \Theta)$ to refer to the expectation operator and the VTS with respect to the pricing measure induced by the Bates model with parameter vector $\Theta$ respectively. Because the augmented objective function in \eqref{eq:obj_function_intro} sums errors along the model-implied VTS, it is useful to have $VTS^{mod}(\tau\vert \Theta)$ in closed form: the objective can then be evaluated exactly for any parameter vector $\Theta$, without computing the integrals in \eqref{eq: VTS} by quadrature. 

\begin{proposition}[Closed-form VTS in the Bates model]\label{prop: bates_vts}
In the Bates model with parameters $\Theta$ and $\kappa > 0$, the VTS at maturity $\tau$ is
\begin{equation}\label{eq: bates_VIX}
VTS^{mod}(\tau\vert \Theta) = \underbrace{\theta+\frac{V_0-\theta}{\kappa \tau}\left(1-e^{-\kappa \tau}\right)}_{\text{average expected variance}} \;+\; \underbrace{2\lambda\left[\mu_J - \log(1+\mu_J) + \frac{\sigma_J^2}{2}\right]}_{\text{jump term}}.
\end{equation}
The first term is the average over $[0,\tau]$ of the expected spot variance, $\mathbb{E}^{\mathbb{Q}}(V_t\vert\Theta)$. The jump term is nonnegative and the same at every maturity.
\end{proposition}

Appendix~\ref{app: vts_proof} proves Proposition~\ref{prop: bates_vts}, which has three consequences for the calibration. First, the model's VTS costs almost nothing to evaluate, so the VTS term in \eqref{eq:obj_function_intro} adds little to the cost of computing the objective. Second, the shape of the model's VTS comes from the variance factor alone. The jump term, which equals $2\lambda\,\mathbb{E}^{\mathbb{Q}}(e^{J} - 1 - J\vert\Theta)$, the jumps' contribution to the log contract in \eqref{eq: VS_cont}, shifts the whole term structure by the same amount. Third, the VTS cannot tell jump laws apart: any two laws with the same jump term give the same VTS, so a penalty on the VTS disciplines the jumps' total contribution to the VTS but not how it is split between the intensity and the jump sizes.

\subsection{Calibration framework}\label{subsec: calibration_framework}
I price options in the Bates model with the COS method of \cite{fang2009novel}, a fast Fourier-based method. The method expands the density of the log return over a truncated interval as a Fourier cosine series whose coefficients follow directly from the model's characteristic function, which is available in closed form for the Bates model.\footnote{For each maturity, I set the interval to $[c_1 - L\sqrt{c_2+\sqrt{c_4}},\; c_1 + L\sqrt{c_2+\sqrt{c_4}}]$, where $c_1$, $c_2$ and $c_4$ are the first, second and fourth cumulants of the log return and $L=12$, widened where necessary to cover the log-moneyness of every listed strike, and I use 512 terms of the expansion.} I compute put prices using this expansion and obtain call prices by put--call parity.

Given a parameter vector $\Theta$, I evaluate \eqref{eq:obj_function_intro} on each date as follows. I price every contract in the date's panel and invert each model price to obtain its \cite{black1976pricing} implied volatility.\footnote{I apply two safeguards to keep the objective finite and continuous in $\Theta$. First, model prices below $10^{-8}$ times the forward are raised to that level before inversion, for numerical stability. Second, during inversion, I consider a range of implied volatility between 0.01\% and 500\%. When no volatility in that range reproduces a model price, which can happen at extreme parameters, I use the nearer end of the range. } I invert market prices to implied volatilities in the same way. I compute $VTS^{mod}(\tau_j\vert\Theta)$ from the closed-form expression \eqref{eq: bates_VIX} and $VTS^{mkt}(\tau_j)$ from the procedure in Section~\ref{subsec: vs}. Finally, I divide \eqref{eq:obj_function_intro} by the number of contracts on the date, $\sum_j N(\tau_j)$. This leaves the minimizer unchanged but makes the objective comparable across dates, as the number of quotes grows from 169 contracts a day on average in 1996--2004 to 5,458 in 2014--2025 (Table~\ref{tab:summary_stats}). 

I minimize the objective with L-BFGS-B over a box of parameter bounds, rescaling each parameter to the unit interval so that finite-difference gradient steps are comparable across parameters of different magnitudes. For the Bates model, I restrict the parameters to the following intervals: $V_0\in[10^{-4},4]$, $\kappa\in[0.01,250]$, $\theta\in[10^{-4},2]$, $\sigma_v\in[0.01,8]$, $\rho\in[-0.999,0.999]$, $\lambda\in[0,100]$, $\mu_J\in[-0.9,1]$ and $\sigma_J\in[10^{-4},1.5]$.\footnote{The bounds are intentionally wide to accommodate the more extreme parameters that emerge on crisis dates.} To reduce the risk of stopping at a poor local minimum, the first fit on each date starts from two points, the previous date's solution and a fixed set of default values, and I keep the one with the smaller objective.\footnote{If the optimizer's first line search fails and it returns its starting point, I restart it from the default values and from two randomly perturbed starts.}

I estimate the model in two ways, and I call each day's estimated model a fit, or fitted model. In the \emph{unrestricted} fits, all eight parameters are fitted on every date. In the \emph{restricted} fits, two pieces of outside information replace parts of the daily fit. First, $V_0$ is pinned on each date $t$ to a high-frequency estimate of diffusive variance, which is the same under the physical and risk-neutral measures:
\begin{equation}\label{eq: v0_pin}
V_{0,t} = 252\left(\frac{1}{5}\sum_{s=t-4}^{t} BV_s + \widehat{ON}_t\right),
\end{equation}
where $BV_t = \frac{\pi}{2}\sum_{i\geq 2}\vert R_{t,i}\vert\,\vert R_{t,i-1}\vert$ is the bipower variation \citep{barndorff2004power} of day $t$'s 5-minute S\&P~500 index log returns $R_{t,i}$, which is robust to jumps, and $\widehat{ON}_t$ is a HAR-type forecast, as in \cite{bollerslev2011tails}, of the mean squared overnight return of the next five trading days.\footnote{The HAR regression is estimated by ordinary least-squares over the underlying's returns.} Second, the jump sizes $(\mu_J,\sigma_J)$ are shared by all dates. I choose them to minimize the sum of the dates' minimized objectives at $\alpha=1/2$ over 75 test dates, 25 spread evenly over each of 1996--2004, 2005--2013 and 2014--2023, with $V_0$ pinned and $\kappa$, $\theta$, $\sigma_v$, $\rho$ and $\lambda$ refitted on each date. The search starts from a grid of multiples of the physical jump sizes, those of the days that a \cite{lee2008jumps} test flags as jumps, and refines the best of them by Nelder--Mead. It converges at $\mu_J \approx -4.2\%$ and $\sigma_J \approx 7.7\%$, and the objective is flat near this optimum. Appendix~\ref{app: pooled} describes the search and its precision. I hold the optimum fixed for every date and every value of $\alpha$; on each date, $\kappa$, $\theta$, $\sigma_v$, $\rho$ and the intensity $\lambda$ remain free. With this shared jump law in hand, I fit the restricted model on the full option panel (described in Section \ref{sec: data}).

\subsection{Hedging with a calibrated model}\label{subsec: hedging_theory}
Section~\ref{subsec: hedging} compares how well the unrestricted and restricted fits of Section~\ref{subsec: calibration_framework} hedge S\&P~500 European-style equity index options. This subsection describes the hedges I evaluate and the theory behind them. The hedges differ in what they take from a fitted model. The Greek hedges use its sensitivities to small moves in the index and its variance. The quadratic hedges also use its jump law, which is where the restricted and unrestricted fits differ most.

\paragraph{Hedging with option Greeks.} Consider an option's price $O_t$. Over a short interval, its value changes with the price of the underlying and with volatility, and a first-order Taylor expansion approximates the change by
\begin{equation}\label{eq: taylor}
dO_t \approx \frac{\partial O}{\partial t}\,dt + \Delta\ dS_t + \mathcal{V}\, d\sigma_t, \qquad \Delta = \frac{\partial O}{\partial S}, \qquad \mathcal{V} = \frac{\partial O}{\partial \sigma},
\end{equation}
where $\sigma_t$ is the volatility that drives the option's price: its implied volatility in the \cite{black1976pricing} model, or the spot variance $V_t$ in the Bates model. The two Greeks, delta and vega, measure the option's exposure to the underlying's price and to its volatility. The underlying and other options on it carry the same two risks in different proportions, so positions in them can offset the option's exposures. A \emph{delta hedge} sells $\Delta$ units of the underlying, which removes the target's first-order exposure to price moves. A \emph{delta--vega hedge} also sells $\mathcal{V}/\mathcal{V}_a$ units of a second option $O_a$, with delta $\Delta_a$ and vega $\mathcal{V}_a$, which removes the first-order exposure to volatility, and it cuts the position in the underlying to $\Delta - (\mathcal{V}/\mathcal{V}_a)\Delta_a$ to offset the second option's own delta. Both hedges are exact only if prices move continuously and the hedge is rebalanced continuously, so that the terms omitted from \eqref{eq: taylor} vanish. In the general case where jumps are present, the market is incomplete and no hedging portfolio of finitely many instruments can be rebalanced to completely offset changes in $O_t$ \citep{cont2003financial}.

I evaluate four hedges based on Greeks. The first two use the \cite{black1976pricing} formula on the forward, with each option at its own implied volatility $\sigma$: $d_1 = [\log(F/K) + \sigma^2\tau/2]/(\sigma\sqrt{\tau})$, with $\Phi$ the standard normal distribution function and $\phi$ its density. The practitioner's \emph{Black--Scholes delta} hedge sells $\Delta^{BS} = e^{-q\tau}\Phi(d_1)$ units of the index for a call and $\Delta^{BS} = e^{-q\tau}[\Phi(d_1)-1]$ for a put. The \emph{Black--Scholes delta--vega} hedge also sells $\mathcal{V}^{BS}/\mathcal{V}^{BS}_a$ units of a second option $O_a$, where $\mathcal{V}^{BS} = e^{-q\tau}S\sqrt{\tau}\,\phi(d_1)$ is the vega of an option at its own implied volatility, and cuts the index position to $\Delta^{BS} - (\mathcal{V}^{BS}/\mathcal{V}^{BS}_a)\Delta^{BS}_a$. The other two use the Greeks of the fitted Bates model. The \emph{model delta} hedge sells the model's delta, $\Delta = \partial O/\partial S$, holding $V_t$ fixed.\footnote{Because the Bates model is homogeneous of degree one in the index level and the strike, its delta at a strike is set by the level and slope of the fitted price curve at that strike \citep{bates2005hedging}. Two fits that match the same smile therefore have nearly the same delta, however different their jump laws.} The \emph{model delta--vega} hedge uses the model's deltas and vegas, where vega is the sensitivity to the spot variance, $\mathcal{V} = \partial O/\partial V$. 

\paragraph{Quadratic hedging.} Quadratic hedging chooses the hedges by minimizing the expected squared hedging error under the pricing measure. Following \cite{follmer1986hedging}, \cite{cont2007hedging} (CTV) choose an initial capital $c$ and a self-financing strategy $\phi_t$ in $m$ traded instruments, the underlying and other options, to minimize
\begin{equation}\label{eq: ctv_obj}
\mathbb{E}^{\mathbb{Q}}\left[\left(c + \int_0^T \phi_t^\top dX_t - Y\right)^2\right],
\end{equation}
where $X_t$ is the vector of the instruments' discounted prices and $Y$ is the target's discounted payoff at its maturity $T$. CTV take the expectation under the pricing measure $\mathbb{Q}$ calibrated to option prices. CTV show that when traded prices are martingales, the problem has a solution under mild conditions, which can fail under an arbitrary measure. 

\begin{theorem}[\cite{cont2007hedging}, Proposition~4]\label{thm: ctv}
Let $W$ be an $n$-dimensional standard Brownian motion and $\tilde{\mathcal{J}}$ a compensated Poisson random measure with intensity $dt\times\nu(dz)$. Let $\mathcal{F}_t$ be the usual associated filtration. Suppose that under $\mathbb{Q}$ the discounted instrument prices and the discounted value of the target, $Y_t = \mathbb{E}^{\mathbb{Q}}[Y\vert\mathcal{F}_t]$, have the representations
\begin{align*}
X_t &= X_0 + \int_0^t B_u\, dW_u + \int_{[0,t]\times\mathbb{R}} \gamma_u(z)\, \tilde{\mathcal{J}}(du\times dz),\\
Y_t &= Y_0 + \int_0^t b_u\, dW_u + \int_{[0,t]\times\mathbb{R}} \gamma^0_u(z)\, \tilde{\mathcal{J}}(du\times dz),
\end{align*}
where $B_t$ is an $m\times n$ matrix, $b_t$ a $1\times n$ vector, $\gamma_t(z)\in\mathbb{R}^m$ and $\gamma^0_t(z)\in\mathbb{R}$, and the coefficients satisfy CTV's regularity conditions. If the matrix
\[
\Omega_t = B_t B_t^\top + \int_{\mathbb{R}} \gamma_t(z)\,\gamma_t(z)^\top\,\nu(dz)
\]
is nonsingular for all $t\in[0,T]$, then \eqref{eq: ctv_obj} is minimized by the initial capital $\hat{c} = \mathbb{E}^{\mathbb{Q}}[Y] = Y_0$, the target's price, and the strategy
\begin{equation}\label{eq: ctv_hedge}
\hat{\phi}_t = \Omega_t^{-1}\left(B_t\, b_t^\top + \int_{\mathbb{R}} \gamma_t(z)\,\gamma^0_t(z)\,\nu(dz)\right).
\end{equation}
\end{theorem}

For any admissible strategy, the expected squared terminal error equals $(c - Y_0)^2$ plus the expected integral over time of the instantaneous variance of the hedged position, because under $\mathbb{Q}$ the hedging error accumulates as a martingale, whose increments over disjoint intervals are uncorrelated. The optimal capital is therefore the target's price, and the optimal strategy minimizes the instantaneous variance at each date separately. That minimization is a regression: $\Omega_t$ is the instruments' instantaneous covariance matrix, the term in parentheses in \eqref{eq: ctv_hedge} is their instantaneous covariance with the target, and $\hat{\phi}_t$ holds the coefficients of a regression, under $\mathbb{Q}$, of the target's instantaneous change in value on those of the instruments. CTV find that hedging with the underlying alone in this way reduces the error of a delta hedge only marginally, whereas combining the underlying with options can reduce the hedging error substantially.

CTV derive explicit hedge ratios mainly for exponential L\'evy models. Proposition~\ref{prop: bates_hedge} applies Theorem~\ref{thm: ctv} to the Bates model, in which the variance is a diffusion correlated with the index. For the value $f(t,S,V)$ of an instrument or a portfolio, write $f_S = \partial f/\partial S$ and $f_V = \partial f/\partial V$, and let
\[
\delta f(J) = f(t, Se^{J}, V) - f(t, S, V)
\]
be the change in $f$ when the index jumps by the log jump $J$. For the index itself, $f_S = 1$, $f_V = 0$ and $\delta f(J) = S(e^{J}-1)$. Conditional expectations, variances and covariances given the information at $t$ are under $\mathbb{Q}$ and are denoted $\mathbb{E}_t$, $\mathrm{Var}_t$ and $\mathrm{Cov}_t$.

\begin{proposition}[Bates optimal quadratic hedging]\label{prop: bates_hedge}
In the Bates model with parameters $\Theta$, let the target have price $C(t,S,V)$, and let the $m$ hedging instruments, the index and $m-1$ options, have prices $X_k(t,S,V)$, $k = 1,\dots,m$, with $X_1 = S$. Suppose that every price is continuously differentiable once in $t$ and twice in $(S,V)$.
\begin{enumerate}
\item[(a)] \emph{Instantaneous covariances.} For any two such prices $f$ and $g$, $\mathrm{Cov}_t(df_t, dg_t) = \omega(f,g)\,dt + o(dt)$, where
\begin{equation}\label{eq: inst_cov}
\begin{aligned}
\omega(f,g) ={}& \underbrace{V S^2 f_S\, g_S + \rho\sigma_v V S\,(f_S\, g_V + f_V\, g_S) + \sigma_v^2 V f_V\, g_V}_{\text{from the two Brownian motions}}\\
&+ \underbrace{\lambda\,\mathbb{E}^{\mathbb{Q}}\big[\delta f(J)\,\delta g(J)\big]}_{\text{from jumps}}.
\end{aligned}
\end{equation}
\item[(b)] \emph{The minimum-variance hedge.} Hold the target long, $\phi_k$ units of each instrument $k$ short, and the balance in cash. Let $\Omega$ be the $m\times m$ matrix with entries $\Omega_{kl} = \omega(X_k, X_l)$, and $\omega_C$ the vector with entries $\omega(C, X_k)$. The instantaneous variance of the position is
\begin{equation}\label{eq: pos_var}
\sigma^2_\Pi(\phi) = \omega(C,C) - 2\,\phi^\top \omega_C + \phi^\top \Omega\, \phi.
\end{equation}
If $\Omega$ is nonsingular, $\sigma^2_\Pi$ has the unique minimizer $\hat{\phi} = \Omega^{-1}\omega_C$, with minimum $\omega(C,C) - \omega_C^\top \Omega^{-1}\omega_C$, and $\hat{\phi}$ is the strategy \eqref{eq: ctv_hedge} of Theorem~\ref{thm: ctv}.
\item[(c)] \emph{The index alone.} With $m = 1$ and $V + \lambda\eta > 0$,
\begin{equation}\label{eq: mv_delta}
\begin{aligned}
\hat{\phi} &= w\left(\Delta + \frac{\rho\sigma_v\mathcal{V}}{S}\right) + (1-w)\,\Delta^J,\\
w &= \frac{V}{V + \lambda\eta}, \qquad \Delta^J = \frac{\mathbb{E}^{\mathbb{Q}}\big[(e^{J}-1)\,\delta C(J)\big]}{S\,\eta},
\end{aligned}
\end{equation}
where $\Delta = C_S$, $\mathcal{V} = C_V$ and $\eta = \mathbb{E}^{\mathbb{Q}}\big[(e^{J}-1)^2\big] = \mu_J^2 + (1+\mu_J)^2\big(e^{\sigma_J^2}-1\big)$.
\item[(d)] \emph{No jumps.} If $\lambda = 0$, $V > 0$, $\sigma_v > 0$ and $\vert\rho\vert < 1$, and the instruments are the index and one option $O$ with vega $\mathcal{V}_O \neq 0$, then $\hat{\phi}$ is the delta--vega hedge, $\hat{\phi}_O = \mathcal{V}/\mathcal{V}_O$ and $\hat{\phi}_S = \Delta - \hat{\phi}_O\Delta_O$, and the hedged position has zero instantaneous variance.
\end{enumerate}
\end{proposition}

Appendix~\ref{app: hedge_proof} proves Proposition~\ref{prop: bates_hedge}. When the quadratic hedge uses only the index, it is a weighted average of two deltas. The \emph{diffusive delta}, $\Delta + \rho\sigma_v\mathcal{V}/S$, is the change in the option's value expected from the variance move that accompanies a small move in the index: with $\rho < 0$, variance tends to rise when the index falls. The \emph{jump delta}, $\Delta^J$, is the slope of a regression of the option's change in value on the index's change across whole jumps, rather than the tangent at the current index level. The weights are the shares of the index's instantaneous variance that come from the diffusion and from jumps, so a jump law with a small intensity but large jumps can carry a large weight. Minimum-variance deltas of this kind have been studied for index options by \cite{poulsen2009risk}, \cite{hull2017optimal} and, in the \cite{heston1993closed} model without jumps, \cite{alexander2012does}.

Without jumps, the quadratic hedge with the index and one option is the delta--vega hedge and is perfect in continuous time. With jumps, the jump terms in $\Omega$ and $\omega_C$ move it away from the delta--vega hedge. The hedge then accepts some exposure to the Brownian motions to reduce its exposure to jumps, in proportion to their variances under the fitted $\mathbb{Q}$. With the index and two options there are three instruments but only two Brownian motions, so some combination of the instruments has no diffusive exposure and is purely exposed to jumps. 

\paragraph{Regularization.} When the instruments are close substitutes, $\Omega$ can be nearly singular and $\Omega^{-1}\omega_C$ unstable; with two options and $\lambda$ close to zero, for example, the combination exposed to jumps alone has almost no variance. CTV (Remark~6) therefore suggest Tikhonov regularization: penalizing the size of the positions with a weight $\zeta > 0$ replaces $\Omega^{-1}\omega_C$ by $(\Omega + \zeta I)^{-1}\omega_C$, with $\zeta$ chosen, following \cite{engl1996regularization}, so that the hedging error of the regularized strategy is at its highest acceptable level. I scale each instrument by its own instantaneous standard deviation before penalizing it, so that the hedge weights solve $(\Omega + \zeta D^2)\phi = \omega_C$ with $D^2 = \mathrm{diag}(\Omega_{11},\dots,\Omega_{mm})$, and I take the largest $\zeta$ on a grid for which the residual variance $\sigma^2_\Pi(\phi)$ exceeds its unregularized minimum by at most 1\%.\footnote{The grid is $\zeta\in\{0, 10^{-8}, 10^{-7.5}, \dots, 10^{-0.5}, 1\}$. The tolerance is 1\% of the larger of the unregularized minimum and $10^{-4}\,\omega(C,C)$, so that it does not vanish when the model hedge is nearly perfect. The scaling and the 1\% tolerance choices are heuristics.} 

\paragraph{Hedging strategies.} I consider three types of hedges: the practitioner's Black--Scholes hedges, which use no fitted model; hedges built from the sensitivities of a fitted model; and quadratic hedges, which use the fitted model's whole risk-neutral law, jumps included (Table~\ref{tab:hedges}). The practitioner's hedges serve as benchmarks: the delta hedge (H1) trades the index alone, and the delta--vega hedge (H2) also holds an option to offset the target's vega. Both evaluate each option's Black--Scholes delta and vega at that option's own implied volatility. Their hedge ratios therefore do not come from a single, consistent pricing process. The Black--Scholes model has one volatility for the diffusion of the index, whereas these hedges use a different volatility for each option, as if the index followed a different diffusion for each one. The model hedges replace the Black--Scholes Greeks with the fitted model's, which all come from one fitted stochastic process. The model delta (H3) trades the index alone, like H1, and the model delta--vega hedge (H4) adds an option, like H2. That option now offsets the target's exposure to the model's spot variance rather than to its own implied volatility. With the index alone (H5), the quadratic hedge adjusts the model delta for the leverage effect and for jumps, through the jump delta of \eqref{eq: mv_delta}. With the index and one option (H6), it trades the same instruments as H2 and H4 but also hedges jump risk; without jumps, it reduces to H4. With the index and two options (H7), it can also hold a combination of the instruments that is exposed to jumps alone. In the tests of Section~\ref{subsec: hedging}, each day's hedging portfolio is constructed using that day's option prices and calibrated parameters. This portfolio is held without rebalancing, either to the next trading day's close or for a week, depending on the hedging horizon.

\begin{table}[!t]
\centering
\footnotesize
\begin{tabularx}{\textwidth}{@{}l>{\raggedright\arraybackslash}Xll>{\raggedright\arraybackslash}X@{}}
\toprule
 & Hedge & Instruments & Positions & Uses from the fit \\
\midrule
H1 & Black--Scholes delta at own IV & index & $\phi_S = \Delta^{BS}$ & nothing \\
H2 & Black--Scholes delta--vega at own IVs & index, $O_a$ & $\phi_a = \mathcal{V}^{BS}/\mathcal{V}^{BS}_a$, $\phi_S = \Delta^{BS} - \phi_a\Delta^{BS}_a$ & nothing \\
H3 & Model delta & index & $\phi_S = \Delta$ & the fitted smile \\
H4 & Model delta--vega & index, $O_a$ & $\phi_a = \mathcal{V}/\mathcal{V}_a$, $\phi_S = \Delta - \phi_a\Delta_a$ & deltas and vegas \\
H5 & Quadratic, index & index & \eqref{eq: mv_delta}, regularized & the whole $\mathbb{Q}$ law \\
H6 & Quadratic, index and one option & index, $O_a$ & $(\Omega + \zeta D^2)^{-1}\omega_C$ & the whole $\mathbb{Q}$ law \\
H7 & Quadratic, index and two options & index, $O_a$, $O_b$ & $(\Omega + \zeta D^2)^{-1}\omega_C$ & the whole $\mathbb{Q}$ law \\
\bottomrule
\end{tabularx}
\caption{Hedging strategies. $O_a$ and $O_b$ are options of the target's own expiry, described in Section~\ref{subsec: hedging}; $\Omega$, $\omega_C$, $D$ and $\zeta$ are defined in Proposition~\ref{prop: bates_hedge} and the regularization paragraph.}
\label{tab:hedges}
\end{table}

\section{S\&P 500 option panel and variance term structure}\label{sec: data}
\subsection{Option panel}\label{subsec: panel}
I use quotes provided by OptionMetrics for European-style call and put options on the S\&P 500 Index (SPX). The sample runs from January 4, 1996 to August 29, 2025, a total of 7,463 trading days. Quotes are end-of-day 4pm bid and ask prices. For model calibration, I use the contract's mid-price, the average of its bid and ask quotes, as the observed market price. The panel includes both the standard AM-settled contracts, which settle at the opening of trading on their expiration date, and PM-settled contracts, which expire at the close and include the weekly and end-of-month expirations.\footnote{Standard expiration U.S. equity index options are AM-settled and typically expire on the third Friday of each month.} I measure time to maturity in calendar days from the 4pm quote to expiration, which for an AM-settled contract is seven hours before the close of its expiration date. For each maturity, I take the forward price $F_\tau$ from OptionMetrics, which reports one for every maturity in the panel, and I discount with a risk-free rate derived from T-bill yields, applied as a flat curve across maturities. I keep only OTM options, calls with strikes at or above the forward and puts with strikes below it, because, by put--call parity, their prices determine the price of a vanilla option at any strike.\footnote{Since the options are European-style, one can recover the price of an in-the-money option from OTM option prices via put–call parity. From an estimation perspective, OTM options also have the advantage of generally being more liquid than the ITM option of the same strike.} I retain all maturities between one day and one year before applying the filters I describe below.

\begin{table}[!t]
\centering
\footnotesize
\setlength{\tabcolsep}{4pt}
\begin{tabular*}{\textwidth}{@{\extracolsep{\fill}}lrrrrrr}
\toprule
\multicolumn{7}{l}{\textit{Panel A: Composition by period}} \\[2pt]
 & Trading & Maturities & Contracts & Contracts & Shortest & PM-settled \\
 & days & per day & per day & per & maturity & maturities \\
 & & & & maturity & (days) & (\%) \\
\midrule
1996--2004 & 2,265 & 6.0 & 169 & 28 & 16.4 & 0 \\
2005--2013 & 2,265 & 10.7 & 521 & 49 & 6.3 & 39 \\
2014--2025 & 2,933 & 36.8 & 5,458 & 148 & 2.1 & 75 \\
All & 7,463 & 19.5 & 2,354 & 121 & 7.7 & 62 \\
\midrule
\multicolumn{7}{l}{\textit{Panel B: Contracts by maturity and moneyness}} \\[2pt]
 & \multicolumn{3}{c}{Calls} & \multicolumn{3}{c}{Puts} \\
\cmidrule(lr){2-4}\cmidrule(lr){5-7}
 & N (000s) & IV (\%) & Spread & N (000s) & IV (\%) & Spread \\
\midrule
\multicolumn{7}{l}{\textit{1--9 days}} \\
\quad ATM & 197.1 & 18.3 & 0.46 & 196.6 & 21.0 & 0.47 \\
\quad OTM & 178.0 & 17.2 & 0.79 & 182.4 & 23.5 & 0.50 \\
\quad Deep OTM & 105.3 & 20.2 & 1.81 & 523.3 & 31.3 & 1.16 \\
\multicolumn{7}{l}{\textit{10--30 days}} \\
\quad ATM & 760.2 & 15.7 & 0.30 & 837.4 & 19.3 & 0.28 \\
\quad OTM & 461.0 & 14.6 & 0.70 & 613.0 & 22.7 & 0.31 \\
\quad Deep OTM & 189.3 & 17.2 & 1.49 & 1,226.3 & 30.8 & 0.73 \\
\multicolumn{7}{l}{\textit{31--365 days}} \\
\quad ATM & 2,685.2 & 15.6 & 0.30 & 3,083.9 & 21.0 & 0.27 \\
\quad OTM & 971.2 & 14.5 & 0.77 & 2,057.7 & 26.1 & 0.33 \\
\quad Deep OTM & 209.9 & 17.4 & 1.49 & 3,091.4 & 34.3 & 0.87 \\
\bottomrule
\end{tabular*}
\caption{Summary statistics of the S\&P~500 option panel, 1996--2025. Panel A reports averages per trading day, except contracts per maturity and the PM-settled share, which are computed over maturities. Panel B reports the number of contracts, their mean implied volatility and their mean bid--ask spread in implied volatility (volatility points), by days to expiration and standardized moneyness: ATM, OTM and deep OTM for $\vert k\vert$ in $[0,1)$, $[1,2)$ and $[2,6]$.}
\label{tab:summary_stats}
\end{table}

I impose standard filters on the option data from the literature \citep[e.g.,][]{constantinides2013puzzle, goyal2022equity}. These filters eliminate quotes that are unusable or obviously erroneous, prices that violate no-arbitrage bounds, and contracts so deeply OTM that their extrinsic value is negligible. Specifically, I exclude quotes with a zero bid, a bid above the ask, or an ask of 10 cents or less. I also drop contracts whose standardized moneyness exceeds 6 in absolute value and contracts whose mid-price cannot be inverted to a BSIV. The standardized moneyness $k_{i,j}$ of an OTM option with time to maturity $\tau_{j}$ and strike price $K_{i,j}$ is
\[
k_{i,j} = \dfrac{\log\left(K_{i,j}/S_t\right)}{VIX(\tau_{j})\sqrt{\tau_{j}}},
\]
where $S_t$ is the spot price of the index and $VIX(\tau_{j})$ is computed from the maturity's quotes that pass the other filters, with the discretization of \eqref{eq: VTS} used for the VIX index.\footnote{With the maturity's OTM strikes $K_1<\dots<K_{n}$, $VIX^2(\tau_j) = \frac{2e^{r\tau_j}}{\tau_j}\big[\sum_{K_i<F_{\tau_j}} P_0(K_i,\tau_j)\,\Delta K_i/K_i^2 + \sum_{K_i\geq F_{\tau_j}} C_0(K_i,\tau_j)\,\Delta K_i/K_i^2\big]$, where $P_0$ and $C_0$ are mid-prices and $\Delta K_i = (K_{i+1}-K_{i-1})/2$, except at the lowest and highest strikes, where $\Delta K_1 = K_2 - K_1$ and $\Delta K_n = K_n - K_{n-1}$. I use this discretization only to standardize moneyness for the filter; the market VTS in the objective is computed from the whole smile as in Section~\ref{subsec: vs}.} Finally, I drop maturities left with fewer than four contracts. Once these filters are applied, each trading day retains at least five maturities. Of the 21.9 million OTM quotes with maturities between one day and one year, the quote filters remove 2.6 million and the moneyness filter 1.7 million, and the remaining filters remove 399 contracts. The final panel has 17.6 million contracts, 5.76 million calls and 11.81 million puts, and consists of 145,731 unique date--maturity pairs.

Table \ref{tab:summary_stats} summarizes the panel. Panel A shows how much its composition changes over the sample. In 1996--2004, a typical day has six maturities, all AM-settled, and 169 contracts, and the shortest maturity is 16 days away on average. With the spread of PM-settled expirations, a typical day in 2014--2025 has 37 maturities and 5,458 contracts. Three quarters of its maturities are PM-settled, and the shortest is about two days away. Contracts per maturity rise from 28 to 148, and the quoted strikes reach further into both tails. Because the VTS term of \eqref{eq:obj_function_intro} weights each maturity by its number of contracts, these changes also shift the weight the objective places across maturities over time.

Panel B splits the contracts by option type, maturity and standardized moneyness. An option is classified as ATM (at-the-money), OTM (out-of-the-money), or deep OTM if the absolute value of its standardized moneyness $\vert k_{i,j} \vert$ is in $[0,1)$, $[1,2)$, or $[2,6]$, respectively. Options are sorted into maturity categories by days to expiration (DTE): 1--9, 10--30, or 31--365 DTE. Within each maturity category, implied volatilities display the familiar smile. For puts, they rise from ATM to deep OTM options. For calls, they dip slightly from ATM to OTM options and are highest for deep OTM options. The smile is most pronounced for puts: deep OTM puts carry the highest BSIVs, 31 to 34\%, reflecting the relatively high cost of hedging large negative jumps \citep{bollerslev2011tails}. Across all maturity categories, OTM puts also have higher BSIVs on average than OTM calls of similar maturity, and this skew is steepest at long maturities. Puts outnumber calls two to one, mostly through deep OTM puts. Most contracts have maturities beyond 30 days. Maturities of at most nine days account for only 8\% of the contracts, despite their recent popularity. They are listed on only a third of the days in 1996--2004, though on every day in 2014--2025. Bid--ask spreads, measured in implied volatility, average about half a volatility point for both calls and puts. They are narrowest for ATM options beyond nine days and widest for deep OTM options, above all for deep OTM calls with at most nine days to expiration. 

\subsection{Variance term structure}\label{subsec: vts_data}
For every trading day and listed maturity in the panel, I compute $VTS^{mkt}(\tau)$ from that maturity's contracts as in Section~\ref{subsec: vs}, for 145,731 date--maturity pairs. Table \ref{tab:vts_stats} reports summary statistics for the market VIX implied by option prices expressed in annualized volatility points. Panel A pools them by days to expiration and Panel B by the level of volatility (as measured by the 30-day VIX). Consistent with prior studies \citep[e.g.,][]{dew2017price, johnson2017risk}, the term structure typically slopes upward and generally its mean and median rise with maturity. Averaged across days, maturities of 31--91 days carry the largest share of the weight in the VTS term of the objective, and maturities of at most nine days the smallest, 6\%. 

\begin{table}[!t]
\centering
\footnotesize
\setlength{\tabcolsep}{4pt}
\begin{tabular*}{\textwidth}{@{\extracolsep{\fill}}lrrrrr}
\toprule
\multicolumn{6}{l}{\textit{Panel A: By maturity}} \\[2pt]
 & \multicolumn{5}{c}{Days to expiration} \\
\cmidrule(lr){2-6}
 & 1--9 & 10--30 & 31--91 & 92--182 & 183--365 \\
\midrule
Date--maturity pairs & 15,778 & 32,555 & 39,234 & 27,462 & 30,702 \\
Mean & 18.0 & 17.7 & 19.1 & 21.4 & 22.4 \\
Standard deviation & 10.3 & 7.8 & 7.1 & 6.5 & 5.7 \\
5th percentile & 8.3 & 9.8 & 11.7 & 14.0 & 15.3 \\
Median & 15.5 & 15.7 & 17.4 & 19.9 & 21.2 \\
95th percentile & 34.6 & 30.9 & 31.5 & 32.7 & 32.3 \\
Contracts per maturity & 88 & 126 & 151 & 140 & 76 \\
Weight in the VTS term (\%) & 6.2 & 17.7 & 35.5 & 18.9 & 21.7 \\
Beyond quoted strikes (\%) & 1.4 & 1.2 & 1.6 & 2.0 & 3.4 \\
Jiang--Tian minus discretized & 0.07 & 0.07 & 0.11 & 0.17 & 0.31 \\
\midrule
\multicolumn{6}{l}{\textit{Panel B: By level of the 30-day VIX}} \\[2pt]
 & Trading & 30-day & 91-day & 182-day & Upward- \\
 & days & VIX & VIX & VIX & sloping (\%) \\
\midrule
Below 15 & 2,242 & 12.7 & 14.6 & 16.1 & 99.0 \\
15--25 & 3,794 & 19.5 & 20.9 & 21.9 & 84.4 \\
25--40 & 1,220 & 29.6 & 29.4 & 29.1 & 48.9 \\
40 and above & 207 & 50.1 & 45.9 & 41.4 & 1.0 \\
All & 7,463 & 19.9 & 21.1 & 21.9 & 80.7 \\
\bottomrule
\end{tabular*}
\caption{Summary statistics of the market variance term structure, 1996--2025, in annualized volatility points ($\sqrt{VTS^{mkt}(\tau)}$, \%). Panel A pools every listed maturity in the panel by days to expiration. Its last three rows report the average share of the day's VTS weight, $N(\tau_j)/\sum_j N(\tau_j)$; the share of $VTS^{mkt}$ from the flat tails beyond the lowest and highest quoted strikes; and the difference from the discretized VIX on the same contracts, in volatility points. Panel B interpolates each day's term structure to 30, 91 and 182 days and groups the days by the 30-day value; upward-sloping days have a higher 182-day than 30-day value.}
\label{tab:vts_stats}
\end{table}

Panel B interpolates each day's term structure to 30, 91 and 182 days, linearly in total variance as CBOE does for the VIX index, and groups the days by the 30-day value.\footnote{On the 2.4\% of days, all before 2011, whose shortest maturity exceeds 30 days, the 30-day value is extrapolated from the two shortest maturities.} The slope of the term structure is linked to the level of volatility, with the slope most likely to be negative when volatility is high. When the 30-day VIX is below 15, as on 30\% of days, the term structure slopes upward on 99\% of them, from 12.7\% at 30 days to 16.1\% at 182 days. Between 25 and 40, it is flat on average and slopes upward on about half the days. Above 40, on 2.8\% of days in my sample, it is inverted on all but 1\% of them, falling from 50.1\% at 30 days to 41.4\% at 182 days. 
 
\begin{figure}[!t]
    \centering
    \includegraphics[width=\linewidth]{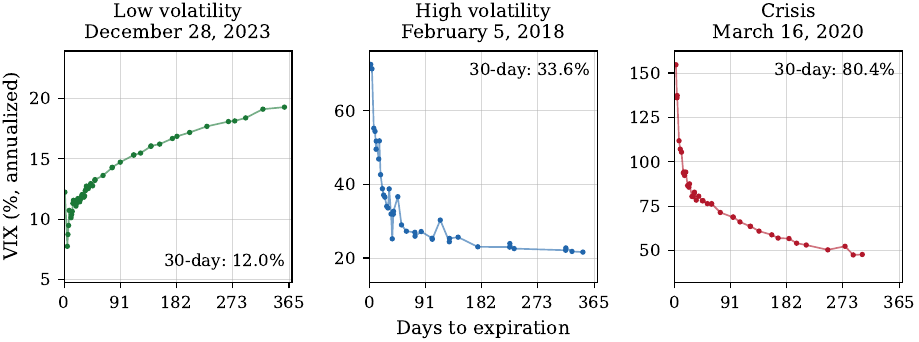}
    \caption{Market VIX, $\sqrt{VTS^{mkt}(\tau)}$, at each listed maturity on three days.}
    \label{fig:vts_days}
\end{figure}

Figure \ref{fig:vts_days} shows the term structure on three days with different levels of volatility. The left plot shows December 28, 2023, a typical low-volatility day, with a 30-day VIX of 12.0\%. The term structure rises from 9.5\% at seven days to 19.3\% at one year. The middle plot shows February 5, 2018, the day dubbed by financial media as ``Volmageddon'' for its sudden spike in volatility. The sudden doubling of the CBOE VIX index sharply raised VIX futures prices and forced the issuers of short-volatility exchange-traded products to buy VIX futures at rising prices, which contributed to the high volatility on the day \citep{augustin2021volmageddon}. The 30-day VIX is 33.6\% and the term structure is inverted, falling from 72.6\% at two days to 21.6\% at 347 days.  The right plot shows March 16, 2020, the trading day following an emergency Federal Reserve rate cut over the prior weekend in response to the COVID-19 pandemic. The S\&P~500 fell sharply after the open, triggering a 15-minute market-wide trading halt, and closed down 12.0\%, while the CBOE VIX ended at a record 82.7. The 30-day VIX is 80.4\%, and the term structure is steeply inverted, falling from 155\% at two days to 48\% near one year.\footnote{The 30-day VIX values in this section are computed from the 4pm quotes as in Section~\ref{subsec: vs}. They differ from CBOE's closing VIX, which is calculated through 4:15pm by a slightly different method. On February 5, 2018, the VIX rose sharply between 4:00 and 4:15pm \citep{augustin2021volmageddon}, so the 4pm quotes understate its closing level.}

The plots show two sources of noise in the term structure. Maturities are measured in calendar days, so a short maturity that spans a weekend or holiday covers fewer trading days, and its annualized variance is lower. On Thursday, December 28, 2023, the five-day maturity expires on Tuesday, January 2, after the New Year weekend and holiday. At 7.8\%, it is the lowest point on the curve, against 12.2\% for the one-day maturity. Another source of noise is maturities with few strikes, whose VTS is estimated imprecisely. On February 5, 2018, the 37-day maturity has only 16 contracts, compared to a median of 98 across all maturities on that day. As the objective in \eqref{eq:obj_function_intro} weights each maturity by its number of contracts, such maturities carry less weight.

\section{Empirical results}\label{sec: emp_results}
This section reports the fits and hedges of the Bates model on the option panel. As described in Section~\ref{subsec: calibration_framework}, I fit the restricted model, with $V_0$ pinned to a high-frequency estimate and one jump-size law for all dates, and the unrestricted model, with all eight parameters free, over different values of $\alpha$. Section~\ref{subsec: model_calibration} compares how closely the fits match the volatility surface and the VTS and how the weight $\alpha$ affects the trade-off in fitting quality between the two. Section~\ref{subsec: hedging} asks whether the fits that match option prices more closely also yield lower hedging errors, with hedges rebalanced daily and weekly, and how the weight $\alpha$ affects hedging performance.

\subsection{Calibration using a VTS penalty}\label{subsec: model_calibration}
To evaluate model fit across the volatility surface and the VTS, Table~\ref{tab:fit_errors} reports mean absolute fitting errors in implied volatility, aggregated across all trading days in the panel: by option type and maturity for the surface in Panel A, and by maturity for the VTS in Panel B. For each set of fits, the column at $\alpha = 1$ is an important special case, as it reduces to the conventional calibration used widely in both academic and industry applications.

\begin{table}[!t]
\centering
\footnotesize
\setlength{\tabcolsep}{4pt}
\begin{tabular*}{\textwidth}{@{\extracolsep{\fill}}lrrrrrrrrr}
\toprule
 & \multicolumn{7}{c}{Restricted, $\alpha =$} & \multicolumn{2}{c}{Unrestricted, $\alpha =$} \\
\cmidrule(lr){2-8}\cmidrule(lr){9-10}
 & 0 & 0.1 & 0.25 & 0.5 & 0.75 & 0.9 & 1 & 0.5 & 1 \\
\midrule
\multicolumn{10}{l}{\textit{Panel A: Implied volatility}} \\[2pt]
Calls, 1--9 days & 2.72 & 2.61 & 2.57 & 2.55 & 2.55 & 2.56 & 2.57 & 1.85 & 1.68 \\
Calls, 10--30 days & 1.75 & 1.63 & 1.59 & 1.56 & 1.56 & 1.56 & 1.57 & 0.85 & 0.62 \\
Calls, 31--365 days & 1.87 & 1.35 & 1.24 & 1.16 & 1.12 & 1.11 & 1.11 & 0.76 & 0.57 \\
Puts, 1--9 days & 2.89 & 2.77 & 2.64 & 2.50 & 2.38 & 2.32 & 2.29 & 1.66 & 1.62 \\
Puts, 10--30 days & 1.55 & 1.49 & 1.45 & 1.40 & 1.36 & 1.35 & 1.35 & 0.69 & 0.65 \\
Puts, 31--365 days & 1.16 & 0.87 & 0.81 & 0.77 & 0.76 & 0.75 & 0.75 & 0.56 & 0.44 \\
All contracts & 1.55 & 1.27 & 1.21 & 1.15 & 1.13 & 1.12 & 1.11 & 0.74 & 0.61 \\
\midrule
\multicolumn{10}{l}{\textit{Panel B: Variance term structure}} \\[2pt]
1--9 days & 1.51 & 1.60 & 1.68 & 1.81 & 1.95 & 2.04 & 2.10 & 1.49 & 2.81 \\
10--30 days & 0.51 & 0.53 & 0.56 & 0.63 & 0.72 & 0.78 & 0.82 & 0.59 & 2.05 \\
31--365 days & 0.29 & 0.29 & 0.31 & 0.35 & 0.41 & 0.45 & 0.49 & 0.30 & 0.74 \\
All maturities & 0.47 & 0.49 & 0.51 & 0.57 & 0.64 & 0.69 & 0.74 & 0.50 & 1.25 \\
\bottomrule
\end{tabular*}
\caption{Mean absolute fitting errors of the Bates model on the S\&P~500 panel, 1996--2025, in volatility points. Panel A pools the implied volatility errors of the OTM options in each bucket; Panel B pools the errors in $\sqrt{VTS}$ at each listed maturity.}
\label{tab:fit_errors}
\end{table}

By maturity, the panels of Table~\ref{tab:fit_errors} show that the shortest maturities are the most difficult for the Bates model to fit. For every set of fits, the implied volatility errors of calls and puts with 1--9 days to expiration are the largest, at 1.62 to 2.89 points, against 0.44 to 1.87 points at longer maturities. The unrestricted fits at $\alpha = 1$ are most directly comparable to those of \cite{broadie2007model}, who estimate the \cite{bates1996jumps} model on S\&P~500 futures options between 1987 and 2003. Calibrating to contracts with 10--180 days to expiration, they report an average error of 0.6 volatility points; the unrestricted fits here have errors of 0.44 to 0.65 points in the 10--30-day and 31--365-day buckets for both calls and puts. The primary difference, other than the slightly different underlying asset, is the composition of maturities in my sample, which can make calibration more challenging. Since 2014, due to the proliferation of short-dated weekday contracts, a typical day has 37 maturities, about half of them PM-settled weekday expirations, and these short maturities must be fitted jointly with longer ones (Table~\ref{tab:summary_stats}). Such near-term maturities are well known to be difficult to fit \citep{bandi2024local, bandi0dte}. \cite{andersen2017short} study options with less than ten days to maturity and find that many parametric option pricing models specify jump processes that are inconsistent with the thickness and time variation of jump tails. Using a Bates-like model, they report an average error of 1.24 volatility points when fitting only their short-maturity sample. The unrestricted fits here have errors of 1.62 to 1.85 points at these maturities, while fitting the entire volatility surface out to one year.

At $\alpha = 1$, the unrestricted fits match the surface closely, with a mean absolute error of 0.61 implied volatility points over all contracts. The VTS errors are relatively large and are largest at the shortest maturities (Table~\ref{tab:fit_errors}, Panel B). The VTS errors are particularly large during crises and grow very large toward the end of the sample, when short-dated maturities are more densely listed. By year, the error stays below one volatility point in every year before 2020 except 2008, but is about two points in every year from 2021 to 2025. Option prices alone do not pin down the VTS, even though the VTS is computed from the same prices. The objective at $\alpha = 1$ weights each contract's implied volatility error equally. The VTS weights the options of a maturity differently, by $1/K^2$ as in \eqref{eq: VTS}, and it weights their prices rather than their implied volatilities. A close fit to the smile in implied volatility therefore need not give a close fit to the VTS. Lowering $\alpha$ to $1/2$ cuts the VTS errors by about half or more at every maturity, and to below one point in each recent year, while the surface errors rise by less than a quarter of a volatility point in any bucket. For the unrestricted fits, then, the VTS penalty sharply reduces the VTS errors at the cost of a small loss in accuracy on the volatility surface.

The restricted fits match the surface less closely, with an error almost twice that of the unrestricted fits at $\alpha = 1$, because pinning $V_0$ and fixing the jump sizes leave fewer parameters to fit each day's smile. Yet at $\alpha = 1$ their VTS errors are lower than those of the unrestricted fits at every maturity. The restrictions discipline the VTS by themselves: by \eqref{eq: bates_VIX}, the model's VTS tends to $V_0$ plus the jump term as the maturity shrinks, so pinning $V_0$ to the high-frequency estimate anchors its short end. \cite{andersen2015parametric} regularize their estimation from option panels in a related way, with a term that penalizes deviations of the model's spot variance from a high-frequency estimate. Lowering $\alpha$ then trades the two fits off along a frontier, and between $\alpha = 1$ and $1/2$ the trade-off is very mild: the VTS errors fall by 14 to 28\%, while the surface error over all contracts rises only from 1.11 to 1.15 points.

\begin{figure}[!t]
    \centering
    \includegraphics[width=0.75\linewidth]{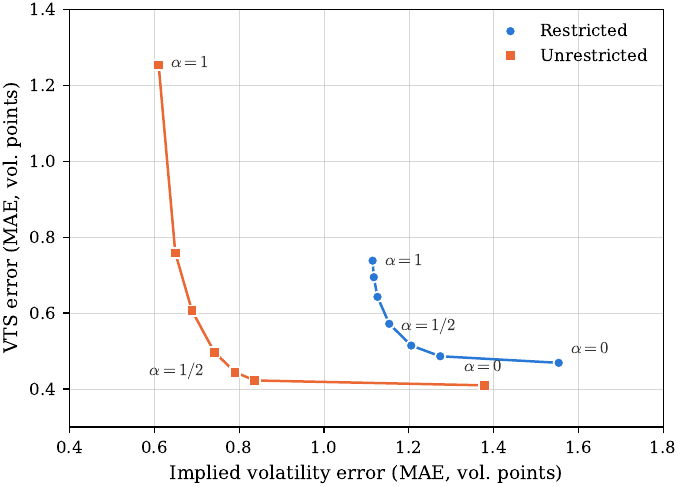}
    \caption{Implied volatility and VTS fitting errors of the restricted and the unrestricted fits as a function of $\alpha$: mean absolute errors over all contracts and over all listed maturities in volatility points.}
    \label{fig:fit_frontier}
\end{figure}

At the other extreme, $\alpha = 0$, the error in contract implied volatilities drops out of the objective function, leaving only information from the VTS to guide calibration. As expected, the restricted fits' VTS error over all maturities is then the lowest in Table~\ref{tab:fit_errors}. Without the implied volatility term, however, the fit to the surface is poorer in every bucket than at any other value of $\alpha$, while the VTS error is only slightly lower than at $\alpha = 1/4$. This is a familiar feature of flexible option pricing models such as the Bates model: many different pricing measures can generate nearly identical variance term structures, so the VTS alone cannot identify the volatility surface.

Figure~\ref{fig:fit_frontier} plots the two errors of each set of fits against each other at the seven values of $\alpha$ in Table~\ref{tab:fit_errors}. Lowering $\alpha$ from 1 to $1/2$ moves the unrestricted fits almost straight down: their VTS error falls by more than half, while their surface error rises by little more than a tenth of a volatility point. Most of that fall comes with the first step, to $\alpha = 0.9$. Below $\alpha = 1/2$ their curve flattens, and at $\alpha = 0$ the surface error rises sharply for almost no further gain on the VTS. The restricted fits trace out a similar frontier, with a minimal trade-off for values of $\alpha$ between $1/2$ and 1. At $\alpha = 1$ and 0.9, the restricted fits have the lower VTS error; at every value of $\alpha$ of 0.75 or less, the unrestricted fits are better on both measures. For both sets of fits, values of $\alpha$ between $1/2$ and 0.9 improve the VTS at little cost in terms of surface error.

\begin{table}[!t]
\centering
\footnotesize
\setlength{\tabcolsep}{2.5pt}
\begin{tabular*}{\textwidth}{@{\extracolsep{\fill}}lrrrr}
\toprule
 & \multicolumn{2}{c}{Restricted} & \multicolumn{2}{c}{Unrestricted} \\
\cmidrule(lr){2-3}\cmidrule(lr){4-5}
 & $\alpha = 1$ & $\alpha = 1/2$ & $\alpha = 1$ & $\alpha = 1/2$ \\
\midrule
$\sqrt{V_0}$ (\%) & 12.5 [8.1, 23.4] & 12.5 [8.1, 23.4] & 16.0 [9.6, 28.5] & 15.3 [8.7, 28.2] \\
$\kappa$ & 5.3 [1.1, 15.1] & 4.9 [1.3, 15.2] & 3.5 [1.0, 10.1] & 4.0 [1.3, 11.8] \\
$\sqrt{\theta}$ (\%) & 20.6 [13.9, 31.2] & 20.2 [12.6, 29.7] & 21.8 [15.6, 32.5] & 21.7 [15.7, 31.4] \\
$\sigma_v$ & 1.63 [0.79, 3.58] & 1.58 [0.77, 3.70] & 0.87 [0.49, 1.84] & 1.01 [0.55, 2.33] \\
$\rho$ & $-$0.70 [$-$0.85, $-$0.55] & $-$0.71 [$-$0.87, $-$0.55] & $-$0.72 [$-$0.84, $-$0.60] & $-$0.70 [$-$0.83, $-$0.58] \\
$\lambda$ & 0.87 [0.09, 4.39] & 1.06 [0.09, 4.73] & 0.06 [0.01, 0.84] & 0.07 [0.01, 1.59] \\
$\mu_J$ (\%) & $-$4.2 (fixed) & $-$4.2 (fixed) & $-$16.4 [$-$53.3, $-$5.3] & $-$14.7 [$-$38.9, $-$3.5] \\
$\sigma_J$ (\%) & 7.7 (fixed) & 7.7 (fixed) & 26.3 [9.3, 70.6] & 21.7 [6.1, 46.3] \\
\midrule
\multicolumn{5}{l}{Days with a mean jump larger than 25\% in absolute value (\%)} \\
 & 0.0 & 0.0 & 30.2 & 25.6 \\
\bottomrule
\end{tabular*}
\caption{Fitted Bates parameters, 1996--2025: median across trading days, with the 10th and 90th percentiles in brackets. The restricted fits pin $V_0$ to the high-frequency estimate \eqref{eq: v0_pin} and fix the jump sizes. The last row is the share of trading days, in percent, whose fitted mean jump $\mu_J$ is larger than 25\% in absolute value.}
\label{tab:params}
\end{table}

Table~\ref{tab:params} reports the fitted parameters. The two sets of fits price options with very different jump laws. The restricted fits see around one jump a year. The shared jump law has a mean of $-4.2\%$ and standard deviation of 7.7\%. The unrestricted fits at $\alpha = 1$ see jumps highly infrequently, around one every 18 years. Their fitted jump laws have a median $\mu_J$ of $-16\%$ and a median $\sigma_J$ of 26\%. Day to day, the jump laws also vary widely. On 30\% of days the mean jump is larger than 25\% in absolute value. Lowering $\alpha$ reduces the extreme jump sizes implied by the fitted parameters, but it does not explicitly exclude such jump laws: the share of days with a mean jump that large falls only to about a quarter, for each value of $\alpha$ considered here between 0.9 and 0.1. The diffusive parameters differ as well. The restricted fits pin the spot variance at the high-frequency estimate, below the unrestricted fits' level, and their variance is more volatile, while the long-run level is similar in both. Despite these large differences in the jump distribution, the two sets of fits come within about half a volatility point of each other on the surface. Option prices alone cannot fully discriminate between a jump law with moderate but more frequent jumps and one with implausibly large jumps which arrive infrequently.

Overall, I draw three conclusions from these fitting exercises. First, option prices alone do not pin down the VTS: the conventional calibration at $\alpha = 1$ fits the surface closely but produces relatively large VTS errors, and a small weight on the VTS sharply reduces the error at the cost of a small loss in accuracy on the volatility surface. Second, the daily fit does not pin down the jump law: very different jump laws fit the surface about equally well, and, left unrestricted, the model can imply an extreme and implausible jump law. Third, the two forms of discipline overlap: the restrictions lower the VTS errors without any penalty, and the penalty reduces the extreme jump sizes implied by the fitted parameters, but it does not explicitly exclude such jump laws.

The unrestricted and restricted fits, and the fits at different values of $\alpha$, therefore deliver different model parameters. Fits that price options comparably well can imply very different responses of option values to a jump, and so different hedging portfolios. Naturally, Section \ref{subsec: hedging} asks whether the fits with lower fitting errors also yield lower hedging errors. I also examine how $\alpha$ affects hedging performance.

\subsection{Hedging performance}\label{subsec: hedging}
On every trading day $t$, I hedge options at five points on the smile: the 10- and 25-delta puts, the at-the-money option, and the 25- and 10-delta calls, each the contract whose \cite{black1976pricing} forward delta at its own implied volatility is nearest the target, within 0.05. I evaluate hedging performance at three maturities: the one closest to 7 days, within 5 to 9 days; the one closest to 30 days, within 14 to 42 days; and the one closest to 91 days, within 60 to 120 days. I refer to them as the 1-week, 1-month and 3-month maturities. Each target is hedged with instruments of its own expiry: the index, that expiry's at-the-money option ($O_a$), and its opposite-side 25-delta option ($O_b$), the 25-delta call for put targets and the 25-delta put for call targets, so that no option is hedged with a near copy of itself.\footnote{The at-the-money target is $O_a$ itself, so it is hedged with its expiry's 25-delta put in place of $O_a$, and with the 25-delta put and call in the two-option hedge. An expiry is skipped on a date if any of its three instruments is missing or has no quote on the next trading day, which removes 801 of 18,555 date--expiry pairs, almost all before 2014. Some days before 2012 have no maturity within 14 to 42 days; they keep their 1-week and 3-month targets.} The hedges of Table~\ref{tab:hedges} are set at the close of day $t$ from that day's fitted models and held to the close of the next trading day $t+1$. Cash positions earn the risk-free rate $r^f_{t}$, and the hedging error of target $i$ is
\begin{equation}\label{eq: hedge_error}
\begin{aligned}
e_{i,t} ={}& \left(O_{i,t+1} - O_{i,t}\right) - \phi_S\left(S_{t+1} - S_t + qS_t\Delta t\right) - \sum_{k\in\{a,b\}} \phi_k\left(O_{k,t+1} - O_{k,t}\right)\\
&- r_{t}^f\Delta t\left(O_{i,t} - \phi_S S_t - \sum_{k\in\{a,b\}} \phi_k O_{k,t}\right),
\end{aligned}
\end{equation}
where $\Delta t$ is the calendar time from $t$ to $t+1$, prices are mid quotes, and errors are expressed in basis points of the index. The unhedged error $u_{i,t}$ sets every position to zero, and a hedge that uses one option or none sets its positions in the unused options to zero. I measure each hedge in two ways: by its root mean squared error and by its hedging effectiveness, $HE = 1 - \sum e_{i,t}^2/\sum u_{i,t}^2$, where the sums run over all targets and dates. HE is the share of the variance of the unhedged P\&L that the hedging portfolio removes. A perfect hedge has $HE = 1$, and a hedge that does worse than not hedging at all has a negative HE. To compare the two sets of fitted models, I apply the \cite{diebold1995comparing} test to the daily difference between their mean squared errors, with Newey--West standard errors \citep{newey1987simple}. The statistic is the mean of that difference divided by its standard error, and it is approximately standard normal when the two sets of fitted models have equal expected squared hedging errors. Positive values mean that the unrestricted fitted models have the larger hedging errors, and values beyond $\pm 1.645$ are significant at the 10 percent level. I also report results separately for the 24 date pairs that end on a day on which the \cite{lee2008jumps} test detects a jump. Of the 7,462 date pairs, 7,445 keep at least one expiry, giving 88,391 targets per set of fitted models.

Following \cite{alexander2012does}, I also hold each hedge for a week, on non-overlapping Wednesdays. Each hedge is set at the close of Wednesday, or of the next trading day of that week if Wednesday is a holiday, and held unchanged to the same day of the following week. The hedging error is \eqref{eq: hedge_error}, with $t+1$ replaced by the same day of the following week and $\Delta t$ the calendar time between the two. The targets and instruments are chosen as for the daily hedges, except that the 1-week maturity must outlive the holding period: it is the expiry closest to 7 days among those at most 9 days away that expire after the end of the holding period. This gives 1,545 weekly date pairs and 18,242 targets per set of fitted models. I refer to a week whose holding period contains a jump day, as detected by the \cite{lee2008jumps} test, as a jump week.

\begin{table}[!t]
\centering
\footnotesize
\setlength{\tabcolsep}{3.5pt}
\begin{tabular*}{\textwidth}{@{\extracolsep{\fill}}lrrrrrrrr}
\toprule
 & \multicolumn{5}{c}{All date pairs} & \multicolumn{3}{c}{Jump days} \\
\cmidrule(lr){2-6}\cmidrule(lr){7-9}
 & \multicolumn{2}{c}{Restricted} & \multicolumn{2}{c}{Unrestricted} & & Restr. & Unrestr. & \\
\cmidrule(lr){2-3}\cmidrule(lr){4-5}
Hedge & HE & RMSE & HE & RMSE & DM & HE & HE & $t$ \\
\midrule
\multicolumn{9}{l}{\textit{Panel A: $\alpha = 1$}} \\[2pt]
Unhedged & --- & 35.0 & --- & 35.0 & --- & --- & --- & --- \\
H1 Black--Scholes delta & 0.875 & 12.4 & 0.875 & 12.4 & --- & 0.785 & 0.785 & --- \\
H2 Black--Scholes delta--vega & 0.943 & 8.3 & 0.943 & 8.3 & --- & 0.958 & 0.958 & --- \\
H3 Model delta & 0.736 & 18.0 & 0.753 & 17.4 & $-9.78$ & 0.578 & 0.607 & $-4.27$ \\
H4 Model delta--vega & 0.917 & 10.1 & 0.931 & 9.2 & $-6.82$ & 0.911 & 0.919 & $-0.89$ \\
H5 Quadratic, index & 0.815 & 15.1 & 0.859 & 13.1 & $-10.59$ & 0.869 & 0.866 & $+0.23$ \\
H6 Quadratic, $+$ATM & 0.934 & 9.0 & 0.928 & 9.4 & $+3.86$ & 0.957 & 0.961 & $-0.01$ \\
H7 Quadratic, $+$ATM$+$25$\Delta$ & 0.893 & 11.4 & 0.808 & 15.3 & $+2.30$ & 0.949 & 0.915 & $+1.13$ \\
\midrule
\multicolumn{9}{l}{\textit{Panel B: $\alpha = 1/2$}} \\[2pt]
Unhedged & --- & 35.0 & --- & 35.0 & --- & --- & --- & --- \\
H1 Black--Scholes delta & 0.875 & 12.4 & 0.875 & 12.4 & --- & 0.785 & 0.785 & --- \\
H2 Black--Scholes delta--vega & 0.943 & 8.3 & 0.943 & 8.3 & --- & 0.958 & 0.958 & --- \\
H3 Model delta & 0.736 & 18.0 & 0.753 & 17.4 & $-8.85$ & 0.575 & 0.606 & $-3.62$ \\
H4 Model delta--vega & 0.916 & 10.2 & 0.929 & 9.3 & $-6.31$ & 0.910 & 0.916 & $-0.68$ \\
H5 Quadratic, index & 0.814 & 15.1 & 0.855 & 13.3 & $-11.03$ & 0.875 & 0.867 & $+0.26$ \\
H6 Quadratic, $+$ATM & 0.932 & 9.1 & 0.931 & 9.2 & $+1.23$ & 0.957 & 0.962 & $-0.03$ \\
H7 Quadratic, $+$ATM$+$25$\Delta$ & 0.891 & 11.5 & 0.811 & 15.2 & $+4.75$ & 0.946 & 0.920 & $+1.07$ \\
\bottomrule
\end{tabular*}
\caption{Daily hedging performance, 1996--2025. HE is hedging effectiveness, one minus the ratio of the sum of squared hedging errors to the sum of squared unhedged errors; RMSE is in basis points of the index. DM is the Diebold--Mariano statistic for the difference between the unrestricted and restricted fits' daily mean squared errors, with Newey--West standard errors (10 lags); $t$ is the plain $t$-statistic for the same difference on the 24 jump days. Negative values favor the unrestricted fits.}
\label{tab:hedging}
\end{table}

Table~\ref{tab:hedging} reports the daily hedging results. The Black--Scholes delta (H1) removes 87\% of the variance of the unhedged P\&L. Adding the at-the-money option with Black--Scholes Greeks (H2) raises this to 94\%, and no hedge built from the fitted models does better over all date pairs. I return to this comparison at the end of this section. Of the five hedges based on a fitted stochastic process, two remove more of the variance than the Black--Scholes delta for both sets of fits: the model delta--vega hedge (H4) and the quadratic hedge with the index and the at-the-money option (H6). The model delta (H3) is the worst of the seven hedges. It holds the spot variance fixed, and so ignores that variance tends to rise when the index falls. The quadratic hedge with the index alone (H5) accounts for this comovement and for jumps: it adds to the model delta a leverage term and the jump delta of Proposition~\ref{prop: bates_hedge}. This closes part, but not all, of the model delta's gap to the Black--Scholes delta. Adding the 25-delta option to the one-option quadratic hedge (H7) lowers HE, to 0.89 for the restricted fits and 0.81 for the unrestricted ones.

Which fit has the lower hedging errors depends on which instruments are available for inclusion in the hedge. With the index alone and with the model delta--vega hedge, the unrestricted fits have lower hedging errors, with Diebold--Mariano statistics of about $-6$ to $-11$ at both values of $\alpha$. Once options enter the quadratic hedge, the restricted fits' hedging errors are as low or lower. With one option, their advantage is significant at $\alpha = 1$ but not at $\alpha = 1/2$. With two options, it is significant at both. The same pattern holds at every positive value of $\alpha$ (Table~\ref{tab:hedging_alpha}).

\begin{figure}[!t]
    \centering
    \includegraphics[width=\linewidth]{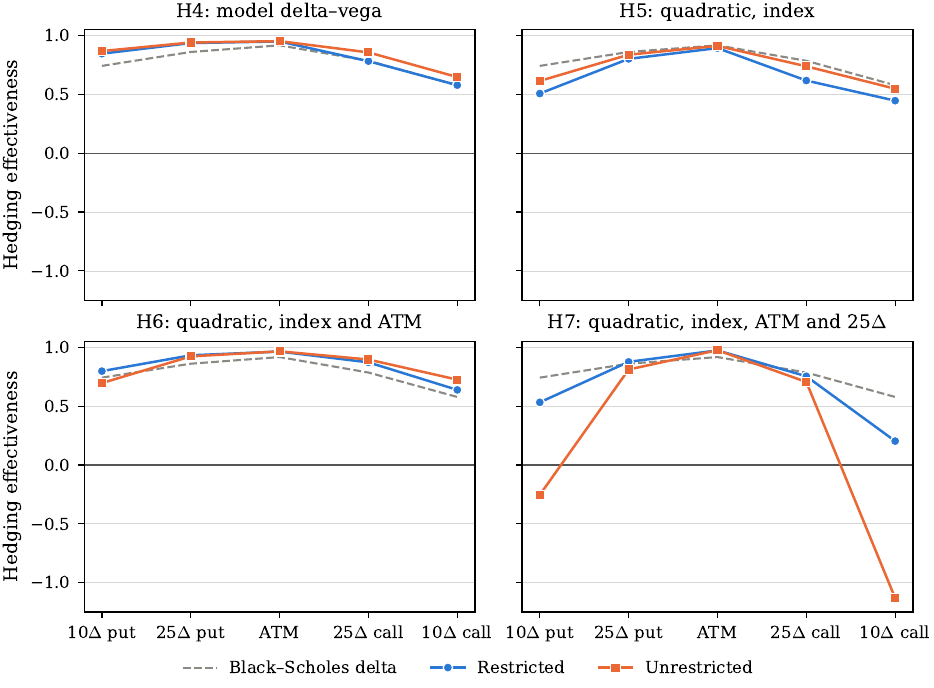}
    \caption{Hedging effectiveness by delta point of the target, pooled over the three maturities, for the restricted and unrestricted fits at $\alpha = 1/2$, 1996--2025, with the Black--Scholes delta for reference. Each panel is one hedge from Table~\ref{tab:hedges}.}
    \label{fig:hedging_smile}
\end{figure}

Figure~\ref{fig:hedging_smile} plots the hedging effectiveness of H4 to H7 at each of the five points of the smile, for both sets of fits at $\alpha = 1/2$. It shows where on the smile the differences between the two sets of fits arise. With the model delta--vega hedge, the unrestricted fits have lower hedging errors at every point of the smile, most of all for the calls. With the quadratic hedge with the index, they also have lower hedging errors at every point. With the index and the at-the-money option, the comparison is mixed: the restricted fits have lower hedging errors for the 10-delta put, and the unrestricted fits for the calls. Adding the second option lowers the effectiveness of both sets of fits' hedges of the 10-delta options, but far less for the restricted fits. Their two-option hedges of these deep out-of-the-money options still remove part of the variance, whereas the unrestricted fits' do worse than not hedging at all. Deep out-of-the-money options are those whose values depend most on the tails of the return distribution, which the jump law shapes. This pattern is therefore consistent with the unrestricted fits' implausible jump laws doing the most damage there. The unrestricted fits' two-option hedges also take larger option positions: the median of the largest absolute option position is 1.26 options per target, against 0.89 for the restricted fits.

Moving from $\alpha = 1$ to $\alpha = 1/2$ changes hedging performance very little, although it raises the fitting errors on the volatility surface (Section~\ref{subsec: model_calibration}). In Tables~\ref{tab:hedging} and~\ref{tab:hedging_weekly}, the hedging effectiveness of every hedge changes by less than 0.01, for either set of fits at either horizon, and every comparison between the two sets of fits keeps its sign. Some of these small changes are statistically significant, and I return to them with Table~\ref{tab:hedging_alpha} below. A small weight on the VTS, then, buys a much better fit to the VTS at almost no cost in hedging.

On the 24 jump days, unhedged errors are almost five times larger than over all date pairs. Hedges based on Greeks work best for small changes in the index and in implied volatility, not for large moves. On jump days, the delta hedges lose the most. The Black--Scholes delta removes 78\% of the variance, against 87\% over all date pairs, and the restricted fits' model delta removes 58\%, against 74\%. The delta--vega hedges, which also trade an option, lose little or nothing. The quadratic hedges, which take positions against jumps, do better on jump days than over all date pairs, with the index alone and with options. With the index and the at-the-money option, they remove about 96\% of the variance, as much as the Black--Scholes delta--vega hedge. The two-option hedges show the effect of imposing a plausible jump law. On these days, the restricted fits' two-option hedge has the higher HE, although with only 24 days the difference is not significant. Otherwise, the two sets of fits' quadratic hedges have comparable hedging errors, and the unrestricted fits have significantly lower hedging errors only with the model delta.

\begin{table}[!t]
\centering
\footnotesize
\setlength{\tabcolsep}{3.5pt}
\begin{tabular*}{\textwidth}{@{\extracolsep{\fill}}lrrrrrrrr}
\toprule
 & \multicolumn{5}{c}{All weeks} & \multicolumn{3}{c}{Jump weeks} \\
\cmidrule(lr){2-6}\cmidrule(lr){7-9}
 & \multicolumn{2}{c}{Restricted} & \multicolumn{2}{c}{Unrestricted} & & Restr. & Unrestr. & \\
\cmidrule(lr){2-3}\cmidrule(lr){4-5}
Hedge & HE & RMSE & HE & RMSE & DM & HE & HE & $t$ \\
\midrule
\multicolumn{9}{l}{\textit{Panel A: $\alpha = 1$}} \\[2pt]
Unhedged & --- & 67.7 & --- & 67.7 & --- & --- & --- & --- \\
H1 Black--Scholes delta & 0.818 & 28.9 & 0.818 & 28.9 & --- & 0.705 & 0.705 & --- \\
H2 Black--Scholes delta--vega & 0.963 & 13.1 & 0.963 & 13.1 & --- & 0.943 & 0.943 & --- \\
H3 Model delta & 0.694 & 37.5 & 0.708 & 36.6 & $-6.78$ & 0.459 & 0.477 & $-1.93$ \\
H4 Model delta--vega & 0.937 & 17.0 & 0.948 & 15.5 & $-1.38$ & 0.928 & 0.877 & $+1.03$ \\
H5 Quadratic, index & 0.745 & 34.2 & 0.790 & 31.0 & $-6.90$ & 0.877 & 0.825 & $+1.27$ \\
H6 Quadratic, $+$ATM & 0.950 & 15.2 & 0.936 & 17.2 & $+3.98$ & 0.955 & 0.961 & $-0.66$ \\
H7 Quadratic, $+$ATM$+$25$\Delta$ & 0.944 & 16.1 & 0.905 & 20.9 & $+4.06$ & 0.959 & 0.944 & $+2.28$ \\
\midrule
\multicolumn{9}{l}{\textit{Panel B: $\alpha = 1/2$}} \\[2pt]
Unhedged & --- & 67.7 & --- & 67.7 & --- & --- & --- & --- \\
H1 Black--Scholes delta & 0.818 & 28.9 & 0.818 & 28.9 & --- & 0.705 & 0.705 & --- \\
H2 Black--Scholes delta--vega & 0.963 & 13.1 & 0.963 & 13.1 & --- & 0.943 & 0.943 & --- \\
H3 Model delta & 0.694 & 37.5 & 0.708 & 36.6 & $-6.29$ & 0.462 & 0.474 & $-0.88$ \\
H4 Model delta--vega & 0.935 & 17.3 & 0.945 & 15.8 & $-1.32$ & 0.927 & 0.870 & $+1.06$ \\
H5 Quadratic, index & 0.742 & 34.4 & 0.789 & 31.1 & $-7.43$ & 0.874 & 0.831 & $+1.08$ \\
H6 Quadratic, $+$ATM & 0.947 & 15.5 & 0.943 & 16.2 & $+1.10$ & 0.954 & 0.967 & $-1.33$ \\
H7 Quadratic, $+$ATM$+$25$\Delta$ & 0.943 & 16.2 & 0.914 & 19.8 & $+7.92$ & 0.956 & 0.944 & $+1.47$ \\
\bottomrule
\end{tabular*}
\caption{Weekly hedging performance, 1996--2025. Each hedge is set at the close of a Wednesday, or of the next trade date of that week, and held unchanged to the same day of the next week. Columns as in Table~\ref{tab:hedging}; the Diebold--Mariano statistics use the weekly mean squared errors, and jump weeks are the weeks that contain a jump day.}
\label{tab:hedging_weekly}
\end{table}

Table~\ref{tab:hedging_weekly} repeats the comparison with hedges held for a week without rebalancing. Unhedged errors are about twice as large as over a day. The hedges that trade the index alone remove less of the variance than daily: the Black--Scholes delta removes 82\%, against 87\%. The hedges that also hold options remove more: the Black--Scholes delta--vega hedge removes 96\%, against 94\%, and the gain is largest for the two-option quadratic hedges. An option offsets the curvature of the target's value in the index, which a position in the index alone cannot. The comparison between the two sets of fits is the same as daily. With the index alone, the unrestricted fits have lower hedging errors, while the model delta--vega hedge no longer separates the two. Once options enter the quadratic hedge, the restricted fits have lower hedging errors, significantly so with two options at both values of $\alpha$ and with one option at $\alpha = 1$. Over a week, the second option costs the restricted fits almost nothing, whereas it slightly lowers the unrestricted fits' effectiveness. The hedge that relies most on the jump law gains most from a plausible one. In the 21 jump weeks, the unrestricted fits have significantly lower hedging errors only with the model delta at $\alpha = 1$, while the restricted fits' two-option hedge has a significant advantage at $\alpha = 1$.

The unrestricted fits' jump laws put the jump risk in rare and implausibly large jumps (Table~\ref{tab:params}). That matters for some hedges and not for others. A delta or delta--vega hedge takes its positions from sensitivities to small moves in the index and its variance, and a fit that matches the smile more closely is likely to measure those sensitivities more accurately. A quadratic hedge with options also takes positions against jumps, sized by the fitted jump law (Proposition~\ref{prop: bates_hedge}). With one option, it accepts some exposure to small moves to reduce its exposure to jumps, and with two, it can hold a combination of options exposed to jumps alone. When calibrated parameters imply an implausible jump law, the unrestricted fits' quadratic hedges are positioned against moves that seldom occur but are large. The results are consistent with this explanation. The unrestricted fits' two-option hedges take larger option positions and fail for deep out-of-the-money options, and their one-option hedge improves as the VTS penalty reduces the extreme jump sizes of their fits (Table~\ref{tab:hedging_alpha}). I find that a plausible jump law matters most for the effectiveness of quadratic hedging with options.

\begin{table}[!t]
\centering
\footnotesize
\setlength{\tabcolsep}{3.5pt}
\begin{tabular*}{\textwidth}{@{\extracolsep{\fill}}lrrrrrrrrrrrr}
\toprule
 & \multicolumn{5}{c}{Restricted} & \multicolumn{5}{c}{Unrestricted} & \multicolumn{2}{c}{DM} \\
\cmidrule(lr){2-6}\cmidrule(lr){7-11}\cmidrule(lr){12-13}
$\alpha$ & H3 & H4 & H5 & H6 & H7 & H3 & H4 & H5 & H6 & H7 & H6 & H7 \\
\midrule
\multicolumn{13}{l}{\textit{Panel A: Daily}} \\[2pt]
0 & 0.734 & 0.869 & 0.814 & 0.927 & 0.879 & 0.743 & $-$13.194 & 0.804 & 0.909 & $-$0.575 & $+3.00$ & $+1.29$ \\
0.1 & 0.734 & 0.887 & 0.814 & 0.930 & 0.887 & 0.753 & 0.926 & 0.848 & 0.930 & 0.805 & $+0.85$ & $+4.60$ \\
0.25 & 0.735 & 0.902 & 0.814 & 0.931 & 0.890 & 0.753 & 0.928 & 0.852 & 0.931 & 0.809 & $+0.57$ & $+4.64$ \\
0.5 & 0.736 & 0.916 & 0.814 & 0.932 & 0.891 & 0.753 & 0.929 & 0.855 & 0.931 & 0.811 & $+1.23$ & $+4.75$ \\
0.75 & 0.737 & 0.917 & 0.815 & 0.933 & 0.893 & 0.753 & 0.930 & 0.856 & 0.931 & 0.808 & $+2.06$ & $+4.99$ \\
0.9 & 0.736 & 0.917 & 0.816 & 0.934 & 0.893 & 0.753 & 0.931 & 0.858 & 0.930 & 0.806 & $+2.75$ & $+4.94$ \\
1 & 0.736 & 0.917 & 0.815 & 0.934 & 0.893 & 0.753 & 0.931 & 0.859 & 0.928 & 0.808 & $+3.86$ & $+2.30$ \\
\midrule
\multicolumn{13}{l}{\textit{Panel B: Weekly}} \\[2pt]
0 & 0.693 & 0.860 & 0.740 & 0.943 & 0.933 & 0.699 & $-$106.538 & 0.747 & 0.922 & 0.862 & $+2.17$ & $+1.75$ \\
0.1 & 0.692 & 0.897 & 0.739 & 0.945 & 0.940 & 0.709 & 0.939 & 0.784 & 0.944 & 0.913 & $+0.10$ & $+7.04$ \\
0.25 & 0.693 & 0.930 & 0.741 & 0.946 & 0.941 & 0.708 & 0.942 & 0.788 & 0.943 & 0.914 & $+0.82$ & $+7.62$ \\
0.5 & 0.694 & 0.935 & 0.742 & 0.947 & 0.943 & 0.708 & 0.945 & 0.789 & 0.943 & 0.914 & $+1.10$ & $+7.92$ \\
0.75 & 0.694 & 0.936 & 0.744 & 0.948 & 0.943 & 0.708 & 0.946 & 0.790 & 0.941 & 0.911 & $+2.00$ & $+7.06$ \\
0.9 & 0.694 & 0.937 & 0.745 & 0.949 & 0.944 & 0.708 & 0.946 & 0.791 & 0.940 & 0.911 & $+2.68$ & $+8.11$ \\
1 & 0.694 & 0.937 & 0.745 & 0.950 & 0.944 & 0.708 & 0.948 & 0.790 & 0.936 & 0.905 & $+3.98$ & $+4.06$ \\
\bottomrule
\end{tabular*}
\caption{Hedging effectiveness of the hedges that use a fitted model, for the restricted and unrestricted fits at every value of $\alpha$, 1996--2025: the model delta (H3), the model delta--vega hedge (H4), and the quadratic hedges with the index (H5), the index and the at-the-money option (H6), and the index and two options (H7), as in Table~\ref{tab:hedges}. Panel A rebalances daily and Panel B weekly, as in Tables~\ref{tab:hedging} and~\ref{tab:hedging_weekly}. DM is the Diebold--Mariano statistic of those tables, for the quadratic hedges with options; positive values favor the restricted fits. The Black--Scholes hedges do not depend on the fit: H1 has HE of 0.875 daily and 0.818 weekly, and H2 0.943 and 0.963. At $\alpha = 0$, the VTS alone does not identify the unrestricted model, and its H4 and H7 entries reflect days on which the fit stops at a bound of the parameter space.}
\label{tab:hedging_alpha}
\end{table}

Table~\ref{tab:hedging_alpha} extends the comparison of $\alpha = 1$ and $\alpha = 1/2$ to all seven values of $\alpha$. It reports the hedging effectiveness of the five model hedges for both sets of fits, daily and weekly, and the Diebold--Mariano statistics of the two quadratic hedges with options. The weight on the VTS matters much less for hedging than the choice between the two sets of fits. With two options, the restricted fits' hedging errors are significantly lower at every positive value of $\alpha$. For hedging, then, any $\alpha$ between $1/2$ and 1 serves about equally well. Section~\ref{subsec: model_calibration} showed that values of $\alpha$ between $1/2$ and 0.9 sharply reduce the VTS errors at the cost of a small loss in accuracy on the surface (Figure~\ref{fig:fit_frontier}). The hedging results show that these values also leave hedging performance nearly unchanged.

The weight does change the size of the restricted fits' advantage once options enter the quadratic hedge. With one option, the advantage narrows as the weight on the VTS rises. From $\alpha = 1/2$ down to 0.1, it is not significant at either horizon. Two changes close the gap. The unrestricted fits' one-option hedge improves significantly between $\alpha = 1$ and $1/2$. This is consistent with the penalty reducing the extreme jump sizes implied by their fitted parameters. Over the same range, the restricted fits' one-option hedge becomes slightly but significantly worse. With two options, the restricted fits' advantage stays significant at every positive value of $\alpha$. The penalty thus brings the unrestricted fits' one-option hedge level with the restricted fits', but not their two-option hedge. Below $\alpha = 1/2$, the restricted fits' model delta--vega hedge weakens. Their other hedges stay within about 0.01 of their values at $\alpha = 1$.

The unrestricted fits at $\alpha = 0$ place full weight on the VTS. Their model delta--vega hedge does worse than not hedging at both horizons, and their two-option quadratic hedge does so daily. Without option prices, the unrestricted model is not identified. By \eqref{eq: bates_VIX}, the VTS depends only on $V_0$, $\kappa$, $\theta$ and the jump term, so many parameter vectors fit it equally well. On some days, the calibration process stops at a bound of the parameter space, which can produce unusual implications. With $\kappa$ at its upper bound, for example, spot variance reverts within days, and option prices respond little to it. The delta--vega hedge, whose option position is a ratio of vegas, then takes extreme positions. The restricted fits, in general, do not imply large positions, and overall they see better hedging performance. This is likely because pinning $V_0$ and fixing the jump sizes leave the VTS fewer parameters to determine.

None of the model hedges improves on the Black--Scholes delta--vega hedge (H2), which removes 94\% of the variance daily and 96\% weekly. Every model hedge with an option has a significantly larger mean squared error, for either fit, either value of $\alpha$ and either horizon. This is not new. \cite{alexander2012does} find the same for the \cite{heston1993closed} model on FTSE~100 options: under every calibration they consider, and with both daily and weekly rebalancing, the practitioner's sticky-strike delta--vega hedge outperforms the model's. The comparison is not like for like. The Black--Scholes model has a single volatility parameter for the diffusion of the index, so as a pricing model it values every option with the same volatility. The practitioner's hedge instead takes each option's Greeks at that option's own implied volatility, as if the index followed a different diffusion for each option. Its hedge ratios therefore rest on what \cite{alexander2012does} call ``deliberately inconsistent assumptions about the price process'', whereas a model hedge takes every option's sensitivities from one fitted process that matches the smile only approximately. A closer fit to the VTS does not help: at $\alpha = 1/2$, the model delta--vega hedge is slightly but significantly worse than at $\alpha = 1$ for both fits. Only in periods with jumps do the quadratic hedges with one option match it, on jump days and in jump weeks alike.

Lower fitting errors on the volatility surface do not necessarily translate to lower hedging errors. The unrestricted fits have lower hedging errors when the hedge trades the index alone or neutralizes first-order exposures to price and variance. Once the quadratic hedge trades options, the restricted fits' hedging errors are as low or lower, daily and weekly, and with two options their advantage is significant at every positive value of $\alpha$. The difference is largest for deep out-of-the-money options, whose values depend most on the jump law. For these options, the unrestricted fits' two-option hedges do worse than not hedging at all. A second option costs the restricted fits much less effectiveness than the unrestricted fits, and over a week almost none. The weight on the VTS, by contrast, changes hedging performance little between $\alpha = 1$ and $1/2$. A calibration can therefore be chosen for a plausible jump law and a close fit to the VTS, at some cost in fit to the surface, and still hedge well with options.

\section{Conclusion}\label{sec: conclusion}
Calibrating an option pricing model is routine in industry practice and academic research, and the conventional objective, the sum of squared errors in implied volatility, rewards the closest fit to option prices. I calibrate the \cite{bates1996jumps} model to S\&P~500 index options on each trading day from 1996 to 2025 with an objective that combines errors on the volatility surface with errors on the VTS, and I vary the relative weight $\alpha$ on the two, from the conventional calibration at $\alpha = 1$ to a fit to the VTS alone at $\alpha = 0$, to trace out the trade-off between the errors on the surface and on the VTS. I fit the model in two ways: unrestricted, with all eight parameters free each day, and restricted, the calibration this article proposes, with the spot variance pinned to a high-frequency estimate and one jump-size law shared across dates. I then ask how these choices affect hedging, with seven hedges, from Black--Scholes deltas to quadratic hedges with options, rebalanced daily and weekly.

I find that option prices by themselves do not pin down the VTS, even though the VTS is computed from those same prices: calibrated the conventional way, the model misses it most at the shortest maturities, in crises and in recent years. I find that a small weight on the VTS is sufficient to sharply reduce VTS errors at the cost of a small loss in accuracy on the volatility surface. For the unrestricted fits, this result is apparent when comparing calibration results using $\alpha = 1$ and $\alpha = 0.9$. Reducing $\alpha$ below $1/2$ produces smaller gains on the VTS and costs increasingly more on the surface. The restrictions discipline the VTS as well, without any penalty, and the penalty reduces the extreme jump sizes implied by the fitted parameters, but it does not explicitly exclude such jump laws: the two forms of discipline overlap but are not substitutes. The penalty needs no data beyond the option panel, since the VTS is computed from the same option prices.

Somewhat surprisingly, I find that the closer fit to option prices is not necessarily the model with superior hedging performance. Which fit has the lower hedging errors depends on which instruments are available for inclusion in the hedge. The unrestricted fits have lower hedging errors when the hedge trades the index alone or neutralizes first-order exposures to price and variance. Once options enter the quadratic hedge, the restricted fits' hedging errors are as low or lower, daily and weekly, and with two options their advantage is significant at every positive value of $\alpha$. The difference is largest for deep out-of-the-money options, for which the unrestricted fits' two-option hedges do worse than not hedging at all. The reason is the extreme distributions implied by the model parameters in the unrestricted fits. Left unrestricted, the daily fit prices options with rare and implausibly large jumps, and a quadratic hedge with options takes positions against jumps sized by that law. The weight on the VTS, by contrast, matters little for hedging: between $\alpha = 1$ and $1/2$, the hedging effectiveness of every hedge changes by less than 0.01. Values of $\alpha$ between $1/2$ and 0.9 therefore improve the calibration to the VTS at the cost of a small loss in accuracy on the surface and at little cost in hedging performance. A calibration can also be chosen for a plausible jump law, at some cost in fit, and still hedge well with options.

None of the model hedges improves on average on the practitioner's Black--Scholes delta--vega hedge, although that comparison is not like for like, since the practitioner's hedge takes each option's Greeks at that option's own implied volatility. The results come from one model, one index and hedges with at most two options, and they leave a broader question open. The \cite{bates1996jumps} model is a jump-diffusion with stochastic volatility, but whole classes of option pricing models do not fit that description, from local volatility models \citep{dupire1994pricing} and pure-jump L\'evy models \citep{madan1998variance, carr2002fine} to rough volatility models \citep{gatheral2020quadratic}. Which of these models benefit from using information from outside the volatility surface, for fitting and for hedging, and to what extent, is a question for further work.\\

\bibliography{joint_calibration}

\begin{appendices}
\section{Proofs}\label{app: proofs}
This appendix proves the two propositions in Section~\ref{sec: theory}.

\subsection{Proof of Proposition~\ref{prop: bates_vts}: the closed-form VTS}\label{app: vts_proof}
Write $\mathbb{E}$ for $\mathbb{E}^{\mathbb{Q}}(\cdot\,\vert\,\Theta)$, and $X_\tau = \log(S_\tau/F_\tau)$ for the log return in excess of the forward. The proof has three parts: the VTS is the price of a claim on $-X_\tau$; the mean of $X_\tau$ follows from the dynamics of $\log S_t$; and the mean of $V_t$ follows from the dynamics of $V_t$.

\begin{proof}
\emph{The VTS is the price of a log contract.} Apply the spanning formula of footnote~\ref{fn: spanning} to the claim that pays $g(S_\tau) = -\log(S_\tau/F_\tau)$ at $\tau$. Since $g(F_\tau) = 0$ and $g''(K) = 1/K^2$, the term in brackets in \eqref{eq: VTS} is the time 0 price of this claim, which is also $e^{-r\tau}\,\mathbb{E}[-X_\tau]$. Hence
\begin{equation}\label{eq: vts_log}
VTS^{mod}(\tau\vert\Theta) = \frac{2}{\tau}\,\mathbb{E}\left[-X_\tau\right].
\end{equation}
This holds in any model, with or without jumps.

\emph{The mean of $X_\tau$.} Between jumps, It\^o's formula gives $d\log S_t = (r - q - \lambda\mu_J - V_t/2)\,dt + \sqrt{V_t}\,dW_t$, and at each jump $\log S_t$ moves by the log jump size. Integrating from 0 to $\tau$ and subtracting $\log(F_\tau/S_0) = (r-q)\tau$,
\[
X_\tau = -\lambda\mu_J\,\tau - \frac{1}{2}\int_0^\tau V_t\,dt + \int_0^\tau \sqrt{V_t}\,dW_t + \sum_{i=1}^{N_\tau} J_i.
\]
The stochastic integral has mean zero, because $V_t$ has finite moments \citep{cox1985theory}. The sum has mean $\lambda\tau\,\mathbb{E}[J]$, because the number of jumps $N_\tau$ has mean $\lambda\tau$ and is independent of their sizes. With \eqref{eq: vts_log},
\begin{equation}\label{eq: log_contract_bates}
VTS^{mod}(\tau\vert\Theta) = \frac{1}{\tau}\int_0^\tau \mathbb{E}[V_t]\,dt + 2\lambda\big(\mu_J - \mathbb{E}[J]\big),\qquad \mathbb{E}[J] = \log(1+\mu_J) - \frac{\sigma_J^2}{2}.
\end{equation}

\emph{The mean of $V_t$.} Taking expectations in $dV_t = \kappa(\theta - V_t)\,dt + \sigma_v\sqrt{V_t}\,dW^v_t$, the mean $m(t) = \mathbb{E}[V_t]$ solves $m'(t) = \kappa\left[\theta - m(t)\right]$ with $m(0) = V_0$, so $m(t) = \theta + (V_0 - \theta)e^{-\kappa t}$. Averaging over $[0,\tau]$,
\begin{equation}\label{eq: mean_var}
\frac{1}{\tau}\int_0^\tau \mathbb{E}[V_t]\,dt = \theta + \frac{V_0 - \theta}{\kappa\tau}\left(1 - e^{-\kappa\tau}\right).
\end{equation}
Substituting \eqref{eq: mean_var} and $\mathbb{E}[J]$ into \eqref{eq: log_contract_bates} gives \eqref{eq: bates_VIX}. The jump term does not depend on $\tau$, and it is nonnegative: since $\mathbb{E}[e^J] = 1 + \mu_J$, it equals $2\lambda\,\mathbb{E}[e^{J} - 1 - J]$, and $e^x \geq 1 + x$ for every real $x$.
\end{proof}

\paragraph{Relation to the variance swap rate.} By \eqref{eq: imp_var} and \eqref{eq: mean_var}, and since $\mathbb{E}[J^2] = (\log(1+\mu_J) - \sigma_J^2/2)^2 + \sigma_J^2$, the VS rate of the Bates model is
\begin{equation}\label{eq: var_swap_bates}
VS^{mod}(\tau\vert \Theta) =\theta+\frac{V_0-\theta}{\kappa \tau}\left(1-e^{-\kappa \tau}\right) + \lambda \left[(\log(1+\mu_J) - \sigma^2_J/2)^2 + \sigma_J^2\right],
\end{equation}
as in \cite{broadie2008effect}. Subtracting \eqref{eq: bates_VIX} gives
\begin{equation}\label{eq: VIX_gap}
VS^{mod}(\tau\vert  \Theta) -  VTS^{mod}(\tau\vert \Theta) = \lambda\,\mathbb{E}[J^2] - 2\lambda\,\mathbb{E}[e^J - 1 - J] = 2\lambda\,\mathbb{E}^\mathbb{Q}(1 + J + J^2/2 - e^J\vert \Theta),
\end{equation}
the gap that \cite{ait2020term} derive and that Section~\ref{subsec: vs} cites. Expanding $e^J$ to third order, the gap is $-(\lambda/3)\,\mathbb{E}[J^3]$ to leading order in the jump size.

\subsection{Proof of Proposition~\ref{prop: bates_hedge}: the Bates quadratic hedge}\label{app: hedge_proof}
\begin{proof}
For a price $f_t = f(t, S_t, V_t)$, It\^o's formula for jump-diffusions \citep{cont2003financial, protter2004stochastic} gives
\begin{equation}\label{eq: ito_jump}
df_t = (\ldots)\,dt + f_S\,S\sqrt{V}\,dW_t + f_V\,\sigma_v\sqrt{V}\,dW^v_t + \delta f(J_t)\,dN_t,
\end{equation}
with every coefficient evaluated just before $t$. The two Brownian terms are the usual ones. The last term is the jump: when $N$ jumps, the index moves from $S$ to $Se^{J_t}$ while $V$ does not move, so $f$ changes by $\delta f(J_t)$. The $dt$ term is known at $t$ and does not enter any covariance. Over a short interval, and up to terms smaller than $dt$, increments multiply as
\begin{gather*}
(dW_t)^2 = (dW^v_t)^2 = dt, \qquad dW_t\,dW^v_t = \rho\,dt,\\
(dN_t)^2 = dN_t, \qquad dW_t\,dN_t = dW^v_t\,dN_t = 0,
\end{gather*}
and $\mathbb{E}_t[\psi(J_t)\,dN_t] = \lambda\,\mathbb{E}^{\mathbb{Q}}[\psi(J)]\,dt$ for a function $\psi$ of the jump size, because a jump arrives with probability $\lambda\,dt$ and its size is independent of its timing.

\emph{Part (a).} Multiplying the random parts of $df_t$ and $dg_t$ with these rules,
\[
df_t\,dg_t = \left[V S^2 f_S\, g_S + \rho\sigma_v V S\,(f_S\, g_V + f_V\, g_S) + \sigma_v^2 V f_V\, g_V\right] dt + \delta f(J_t)\,\delta g(J_t)\,dN_t,
\]
whose expectation is $\omega(f,g)\,dt$. The covariance also subtracts the product of the two random parts' means, $\lambda\,\mathbb{E}^{\mathbb{Q}}[\delta f(J)]\,dt$ and $\lambda\,\mathbb{E}^{\mathbb{Q}}[\delta g(J)]\,dt$, which is of order $(dt)^2$. Hence $\mathrm{Cov}_t(df_t, dg_t) = \omega(f,g)\,dt + o(dt)$.

\emph{Part (b).} Over $[t, t+dt]$, the position gains $dC_t - \sum_k \phi_k\,dX_{k,t}$, plus the dividends on the index and the interest on the cash, which are known at $t$. Its variance is the variance of a linear combination of price changes, so by part (a)
\begin{align*}
\mathrm{Var}_t(d\Pi_t) &= \Big[\omega(C,C) - 2\sum_k \phi_k\,\omega(C,X_k) + \sum_{k,l}\phi_k\phi_l\,\omega(X_k,X_l)\Big]\,dt + o(dt)\\
&= \sigma^2_\Pi(\phi)\,dt + o(dt).
\end{align*}
In the same way, for any vector $x$, $x^\top\Omega x\,dt$ is, up to $o(dt)$, the variance of the change in a portfolio holding $x_k$ units of each instrument, so $x^\top\Omega x \geq 0$: $\Omega$ is positive semidefinite, and $\sigma^2_\Pi$ is a convex quadratic function of $\phi$. Its gradient, $2(\Omega\phi - \omega_C)$, is zero at $\hat{\phi} = \Omega^{-1}\omega_C$. A nonsingular positive semidefinite matrix is positive definite, so $\hat{\phi}$ is the unique minimizer. Substituting it into \eqref{eq: pos_var} gives the minimum, $\omega(C,C) - \omega_C^\top\Omega^{-1}\omega_C$.

To see that $\hat{\phi}$ is the strategy of Theorem~\ref{thm: ctv}, note that in the theorem $B_t\,dW_t + \int\gamma_t(z)\,\tilde{\mathcal{J}}(dt\times dz)$ is the change in the discounted instrument prices. Their conditional covariance matrix is therefore $\Omega_t\,dt$, and their conditional covariance with the discounted target is the term in parentheses in \eqref{eq: ctv_hedge}, times $dt$. Discounting multiplies each change by $e^{-rt}$, up to terms known at $t$; for the index, the discounted gain includes its dividends. By part (a), these covariances are $e^{-2rt}\,\Omega\,dt$ and $e^{-2rt}\,\omega_C\,dt$, and the factor $e^{-2rt}$ cancels in \eqref{eq: ctv_hedge}, which gives $\hat{\phi}_t = \Omega^{-1}\omega_C$.

\emph{Part (c).} With the index alone, $X_1 = S$, whose derivatives are 1 and 0 and whose jump change is $S(e^J - 1)$. Part (a) gives
\begin{align*}
\Omega &= \omega(S,S) = S^2(V + \lambda\eta),\\
\omega_C &= \omega(C,S) = VS^2\Delta + \rho\sigma_v VS\,\mathcal{V} + \lambda S\,\mathbb{E}^{\mathbb{Q}}\big[(e^{J}-1)\,\delta C(J)\big],
\end{align*}
and, since $\Omega > 0$ when $V + \lambda\eta > 0$, dividing,
\[
\hat{\phi} = \frac{\omega_C}{\Omega} = \frac{V}{V + \lambda\eta}\left(\Delta + \frac{\rho\sigma_v\mathcal{V}}{S}\right) + \frac{\lambda\eta}{V + \lambda\eta}\cdot\frac{\mathbb{E}^{\mathbb{Q}}\big[(e^{J}-1)\,\delta C(J)\big]}{S\eta},
\]
which is \eqref{eq: mv_delta}. For $\eta$, the normal moment generating function, $\mathbb{E}^{\mathbb{Q}}[e^{sJ}] = e^{sa + s^2\sigma_J^2/2}$ with $a = \log(1+\mu_J) - \sigma_J^2/2$, gives $\mathbb{E}^{\mathbb{Q}}[e^{J}] = 1+\mu_J$ and $\mathbb{E}^{\mathbb{Q}}[e^{2J}] = (1+\mu_J)^2 e^{\sigma_J^2}$. So $\eta = \mathbb{E}^{\mathbb{Q}}[e^{2J}] - 2\,\mathbb{E}^{\mathbb{Q}}[e^{J}] + 1 = \mu_J^2 + (1+\mu_J)^2(e^{\sigma_J^2}-1)$.

\emph{Part (d).} With $\lambda = 0$, by \eqref{eq: ito_jump} the random part of the position's change is
\[
\left(\Delta - \phi_S - \phi_O\Delta_O\right) S\sqrt{V}\,dW_t + \left(\mathcal{V} - \phi_O\mathcal{V}_O\right) \sigma_v\sqrt{V}\,dW^v_t.
\]
Setting both coefficients to zero gives the delta--vega hedge, at which the position has no random part and so zero variance, the smallest possible. It is the only minimizer because $\Omega$ is nonsingular: by part (a), for the index and the option,
\[
\det\Omega = \omega(S,S)\,\omega(O,O) - \omega(S,O)^2 = (1-\rho^2)\,\sigma_v^2\,V^2 S^2\,\mathcal{V}_O^2 > 0.
\]
\end{proof}

\section{The pooled jump-size law}\label{app: pooled}
The restricted fits of Section~\ref{subsec: calibration_framework} share one jump-size law, $(\mu_J, \sigma_J)$, across all dates, while each date keeps its own intensity $\lambda$. On a single date, option prices identify the jump parameters poorly: they pin down the product of the jump intensity and the mean jump size better than either one \citep{broadie2007model}, so a daily fit can trade frequent small jumps for rare large ones at little cost. Pooling dates lets the jump sizes be determined by many dates' option prices at once. I choose the law to minimize
\begin{equation}\label{eq: pooled_F}
F(\mu_J, \sigma_J) = \sum_{t\in\mathcal{T}}\ \min_{\kappa,\,\theta,\,\sigma_v,\,\rho,\,\lambda}\ L_t(\kappa, \theta, \sigma_v, \rho, \lambda;\, \mu_J, \sigma_J),
\end{equation}
where $\mathcal{T}$ is a set of test dates and $L_t$ is the objective \eqref{eq:obj_function_intro} on date $t$ at $\alpha = 1/2$, divided by the date's number of contracts as in Section~\ref{subsec: calibration_framework}, with $V_0$ pinned by \eqref{eq: v0_pin}. Minimizing $F$ is the same as minimizing the sum of the dates' objectives jointly over the shared jump sizes and every date's own parameters. Because each $L_t$ is an average over the date's contracts, every date carries the same weight, however large its panel.

Each evaluation of $F$ requires a restricted fit on every date in $\mathcal{T}$, so I estimate the law on a sample of 75 dates rather than on the whole panel: 25 trading dates evenly spaced over each of 1996--2004, 2005--2013 and 2014--2023, including the first and last trading date of each period. Each period therefore carries the same weight in $F$, and the sample includes calm periods as well as crisis dates such as October 1, 2008 and April 1, 2020. Dates in 2024 and 2025 are not used. On each date, the fit at a candidate law is run from several starting points, and I keep the one with the smallest objective. Two starts are the date's solutions at the two nearest laws already fitted, measured in $(\log\vert\mu_J\vert, \log\sigma_J)$, with the intensity rescaled so that the jump term of \eqref{eq: bates_VIX}, $2\lambda[\mu_J - \log(1+\mu_J) + \sigma_J^2/2]$, is the same at the new law. A third fits the model's VTS alone to the market VTS: with $V_0$ fixed, it chooses $\kappa$, $\theta$ and the level of the jump term, which requires no option prices, and converts that level into an intensity at the candidate law. Before any law has been fitted, the model's default starting values are used as well.

I minimize $F$ in three steps. First, I evaluate it on a grid anchored at the physical jump sizes. A \cite{lee2008jumps} test applied to daily S\&P~500 returns over 1996--2023 flags the days with jumps, and the model's lognormal law fitted to the returns on those days has, rounded to a tenth of a percentage point, a mean jump of $-3.8\%$ and a jump volatility of $2.8\%$. Multiplying each by 1, 2 and 4 gives a $3\times3$ grid of nine laws, from $(-3.8\%, 2.8\%)$ to $(-15.2\%, 11.2\%)$. Only multiples of at least one are used, because option prices imply larger jumps than index returns do: estimates of jump risk premia from S\&P~500 options put the risk-neutral mean jump well below the physical one \citep{pan2002jump, broadie2007model}. I evaluate the grid outward from its center, so that every law after the first can start from the fits at a neighboring law. Second, the two best laws on the grid, $(-3.8\%, 5.6\%)$ and $(-7.6\%, 11.2\%)$, have the same ratio of jump volatility to mean jump, and I evaluate $F$ at their geometric midpoint, $(-5.37\%, 7.92\%)$, starting from the fits at the two. Third, from the midpoint, a Nelder--Mead search runs in the coordinates $x = (\log\vert\mu_J\vert,\ \log(\sigma_J/\vert\mu_J\vert))$. The first coordinate scales every jump at a fixed ratio of jump volatility to mean jump, the direction along which the two best grid laws lie; the second changes the ratio. The first simplex adds 0.35 to each coordinate in turn. The search stops when its vertices agree to within 0.02 in each coordinate, about 2\% in each jump size, and their values of $F$ agree to within 0.2\% of $F$ at the midpoint, or after 30 evaluations of $F$. It then restarts once from its answer with a simplex of side 0.1 and at most 15 evaluations, so that a stop short of the minimum on a flat surface would show up as a further move. The search converges at $\mu_J = -4.22\%$ and $\sigma_J = 7.66\%$.

$F$ is flat near the optimum: laws with $\mu_J$ between about $-4.1$ and $-4.4\%$ and $\sigma_J$ between about 7.5 and 7.8\% raise it by less than 0.05\%. This flatness is of the same order as the search's tolerance in each coordinate, so the main text reports the law to two significant digits, as $-4.2\%$ and $7.7\%$. Every fit on every date is stored, so a law is never fitted twice and an interrupted search resumes where it stopped; rerunning the whole procedure from the raw data with the replication package returns exactly the same law.

The pooled law has about the same mean jump as the jumps the physical screen flags, $-4.2\%$ against $-3.8\%$, but almost three times their volatility, 7.7\% against 2.8\%. The jumps priced in options are therefore far more dispersed than those found in realized returns. A threshold test misses small jumps, so the screen's sizes describe large returns only and are not a full estimate of the physical law; they serve here as an anchor for the grid, not as a restriction on the fits.

\end{appendices}

\end{document}